\documentclass[preprint]{JHEP3} 
    
\JHEPspecialurl{http://jhep.sissa.it/JOURNAL/JHEP3.tar.gz}   
    
\usepackage{epsfig,multicol}   
    
\newcommand{\beq}{\begin{equation}}

\newcommand{\eeq}{\end{equation}}

\newcommand{\beqar}[1]{\begin{eqnarray}\label{#1}}

\newcommand{\eeqar}{\end{eqnarray}}

\newcommand{\D}{\partial}

\newcommand{\gsla}{\gamma\hspace{-.5em/\hspace{.1em}}}

\newcommand{\esla}{\epsilon\hspace{-.5em/\hspace{.1em}}}

\newcommand{\psla}{p\hspace{-.5em/\hspace{.1em}}}

\newcommand{\ksla}{k\hspace{-.5em/\hspace{.1em}}}

\relax

\title{    
{\Large \bf Quark-Antiquark Exchange in $\gamma^*\,\gamma^*$ Scattering}}    
\author{ J. ~Bartels \\    
II. Institut f\"{u}r Theoretische Physik, Universit\"{a}t     
Hamburg \\    
 Luruper Chaussee 149, 22761 Hamburg, Germany\\   
E-mail: \email{bartels@mail.desy.de} }   
\author{M.~Lublinsky  \\ DESY Theory Group, DESY \\    
 Notkestr. 85, 22607 Hamburg, Germany \\   
E-mail: \email{lublinm@mail.desy.de}}

\preprint{DESY-03-105}    
    
\abstract{    
We calculate the high energy behavior of quark-antiquark exchange in    
$\gamma^*\,\gamma^*$ elastic scattering by summing, to all orders, the     
leading double logarithmic contributions of the QCD ladder diagrams.    
Motivation comes from the LEP data for $\sigma_{tot}^{\gamma^* \,\gamma^*}$    
which indicate the need for secondary reggeon exchange. We show that,    
for large photon virtualities, this exchange is calculable in pQCD.    
This applies, in particular, to parts of the LEP kinematic region.}       
   
\begin{document}

\section{Introduction}

The $\gamma^*\,\gamma^*$ collision at high energies is the unique    
laboratory for testing asymptotic properties of perturbative QCD.    
The virtuality of the photons justifies the use of perturbative     
QCD, and modern electron positron colliders (LEPII, a future linear collider  
NLC)    
allow to measure the total cross section of $\gamma^*\,\gamma^*$ scattering     
at energies where asymptotic predictions of perturbative QCD can be     
expected to set in. The dominant contribution to the process is given by the     
BFKL Pomeron \cite{BFKL} which gives rise to a cross section strongly     
rising with energy $\sigma_{total}^{\gamma^*\,\gamma^*}\sim s^{\alpha_P(0)}$.    
Here $\alpha_P(0)$ is the Pomeron intercept which, in leading order and     
for realistic values of the photon virtualities, lies in the region     
$\alpha_P(0)\simeq 0.3\,-\,0.5$.

There is little doubt that the pomeron will dominate at very high energies,     
and it is expected to be a main contribution at any future linear collider.     
At present, however, the only source    
for experimental data on photon photon collisions is LEP \cite{LEP1,LEP2}.    
These data are at energies which cannot be    
considered as asymptotically large, and it has become clear that at LEP     
energies, the cross section is not yet dominated by the pomeron     
~\cite{BHS,BRL,DDR,MK,NSZ,BFKLP}.     
The data rather indicate the necessity to include, in the theoretical     
description, several corrections. Perturbative corrections     
are due to the quark box (often referred to as QPM contribution);     
recently~\cite{DelDuca} first $O(\alpha_s)$ corrections to the quark box     
have been considered.     
NLO corrections to the BFKL Pomeron are not yet fully available:    
whereas the NLO corrections to the BFKL kernel have been completed    
~\cite{FL}, those     
of the photon impact factor are not yet finished   
~\cite{NLOImpact1,NLOImpact2}.    
Nonperturbative contributions include the soft Pomeron (in the low-$Q^2$     
region) and the exchange of secondary reggeons: the exchange of     
$f_0$ (flavor singlet) or $A_0$, $A_2$ (flavor nonsinglet)    
~\cite{DDR,MK,NSZ}.     
In this paper we address the latter contributions, the corrections     
due to secondary reggeons.    
    
In hadron hadron scattering, secondary reggeons denote the exchange     
of mesons and are of nonperturbative nature. This may change, however, if we     
replace one of the hadrons (or both) by a virtual photon. We know from      
the $\gamma^*p$ process in deep inelastic electron proton scattering     
that the rise of the total cross section with energy becomes substantially     
steeper than in proton proton scattering; this observation has lead to the     
notion of a 'hard Pomeron'. What makes this hard Pomeron particularly     
attractive is that its energy dependence becomes calculable in pQCD.     
In $\gamma^*\gamma^*$ scattering, the high energy behavior should be described     
by the BFKL Pomeron, i.e. pQCD provides an absolute prediction for this     
process.     
In analogy with this, one might expect that also the secondary exchange may     
become accessible to a perturbative analysis, if we replace one or both     
external hadrons by virtual photons. If so, meson exchange will be modelled     
by the exchange of $q \bar{q}$ ladders~\cite{RS},     
and its prediction for the energy dependence     
may be tested in the corrections to BFKL exchange in $\gamma^*\gamma^*$     
scattering. It is the purpose of this paper (and a forthcoming one)     
to investigate this possibility more closely. As a start, we present a    
perturbative QCD calculation of the even-signature flavor nonsinglet    
contribution to the total $\gamma^*\,\gamma^*$ cross section. This    
is done by summing over gluon ladders with two quarks in the $t$-channel.    
The flavor singlet case appears to be more complicated since it involves     
the mixing between quark-antiquark and two-gluon states in the t-channel    
(where the two-gluon state has a helicity content different from the    
so-called     
nonsense-helicity configuration in the BFKL Pomeron). The flavor singlet     
exchange is expected to be somewhat larger than the nonsinglet one     
\cite{BER2}, and we will come back to it in a forthcoming paper.     
    
One of the striking differences between gluon exchange in the BFKL     
calculations and quark-antiquark exchange is the appearance of double     
logarithms~\cite{GGL,KiLi}. As result of this, the intercept of the     
$q\bar{q}$-system is of the order     
$\omega_0^{q\bar{q}}\,=\,\sqrt{const\;\alpha_s}$ (as opposed to     
$\omega_0^{BFKL}\,=\,const\;\alpha_s$    
in the single logarithmic high energy behavior of the BFKL Pomeron),    
and its numerical value can be expected to be large. In fact,        
for $q \bar{q}$ scattering it is known ~\cite{RS} that     
the cross section goes as $\sim s^{\omega_0\,-\,1}$ with     
$\omega_0\,=\,\sqrt{2\,\alpha_s\,C_F/\pi}\simeq 0.5$.     
It is remarkable that this intercept obtained in pQCD is very    
close to the nonperturbative one known from phenomenology.     
For $\gamma^*\,\gamma^*$ scattering we obtain the same result for the    
intercept.     
    
Another crucial feature of the double logarithmic calculation     
is its dependence on the infrared region. In principle,         
the ladder graphs to be summed are infrared safe. However, quite in analogy     
with the BFKL approximation, the contribution from small momenta     
is not believable, and one has to introduce a cutoff scale,     
$\mu_0 \le 1$GeV, which    
separates the infrared from the large momentum region.       
Taking the virtualities of the external photons to be sufficiently large and     
then considering     
the high energy limit one observes that, for not too large energies,      
the result is independent of this infrared cutoff: a natural infrared     
cutoff for the transverse momenta inside the ladder     
appears, which is of the order $Q^4/s$ and decreases with energy.     
Thus with increasing energy this cutoff eventually    
reaches the scale $\mu_0^2$,    
and the integration starts to penetrate the nonperturbative domain.     
In this region    
perturbative QCD cannot be trusted any more. Therefore, for energies $s$     
larger than $Q^4/\mu_0^2$ one has to limit the transverse    
momenta to be larger     
than $\mu_0$, and the perturbative high energy behavior of the $q\bar{q}$     
exchange starts to depend upon the cutoff. This dependence turns out to be     
rather weak: it mainly resides in the pre-exponent and not in the     
exponent of the energy. In our calculation we will study this dependence in     
detail: we will compute the sum of double logarithms

\begin{equation}\label{1.1}    
\sum_{n\geq 1} a_n \alpha_s^{n-1} \left( \ln^2 \frac{s}{Q^2} \right)^n.       
\end{equation}    
When $\frac{Q^4}{s}$ reaches the cutoff $\mu_0^2$ another large     
logarithm appears, $\ln Q^2/\mu_0^2$, which we will include into our analysis:    
\beq\label{1.2}    
\sum_{n \geq 1} \alpha_s^{n-1} \left( a_{n0} \ln ^{2n} \frac{s}{Q^2} + a_{n1}     
\ln ^{2n-1} \frac{s}{Q^2} \ln \frac{Q^2}{\mu_0^2} +...+     
a_{nn} \ln ^{2n} \frac{Q^2}{\mu_0^2} \right).    
\eeq    
In particular we will study the role     
of $\mu_0^2$ in the determination of the pre-exponential behavior of the     
high energy asymptotics.    
    
The task of the double log resummation in $\gamma^*\,\gamma^*$    
collision can be attacked by several methods. First, we follow the     
original paper    
Ref. \cite{GGL} in which the question of double log resummation     
$e^+e^-$ annihilation was addressed first. By direct summation of    
the ladder graphs this method leads to a     
Bethe-Salpeter type equation for the amplitude. Compared to the    
original work of Ref. \cite{GGL} the case of $\gamma^*\,\gamma^*$    
is slightly more complicated, since the additional scale $Q^2$ is involved.     
This complication results in a more sophisticated structure of the    
solution. We apply this method to ladder diagrams which are the     
relevant diagrams for the even-signature exchange. One of our main results      
is the discovery of the 'hard region' $\mu_0^2 < Q^4/s$ in which pQCD     
is reliable.

The second method uses the infrared evolution equation (IREE) for    
the amplitude in the Mellin space. IREE was first derived in    
Ref. \cite{KiLi} in application to quark-quark scattering, and, more recently,    
has been applied  to the calculation of the small-x behavior of    
flavor nonsinglet structure functions  in Ref.~\cite{Ry}  and    
to polarized     
structure functions in Refs. ~\cite{BER2,BER1}.   Within    
the double log accuracy IREE traces the dependence of the amplitude on    
the infrared cutoff $\mu$ of the quark transverse momentum in the    
ladder. The scale $\mu$ is auxiliary in this method. We will see,    
however, that when identified with the real scale of nonperturbative physics,     
$\mu_0^2$, it leads to the same answer as the linear     
equation of the first approach. Use of IREE allows to discuss not only     
the even signature case (with ladder diagrams), but also the   
odd signature case (with non-ladder diagrams).   
    
A third way of handling the quark-antiquark exchange has been described    
in ~\cite{KK,Kirschner}: using the partial wave formalism and the notion of   
the reggeon Green's function, a linear equation     
\'{a} la BFKL has been found for the fermion-antifermion scattering amplitude.     
In \cite{Kirschner} the kernel has been shown to have the same     
conformal properties as the BFKL    
kernel. However, the application of this formulation to          
our problem is not completely straightforward: the coupling of the     
quark-antiquark     
ladder to the external photons involves a careful handling of the    
double logarithmic integrations. We will show how our result for the    
hard region can be obtained from the reggeon Greens's function.

The structure of the paper is following. In    
the next Section (2) notations are introduced, and the quark box    
diagram is discussed. In Section 3 we sum the gluon ladder and    
derive a Bethe-Salpeter type equation for the amplitude. This    
equation is solved in Section 4. We also discuss the solution and   
present the  high energy asymptotics.     
In Section 5 we rederive the same result using the method of the infrared    
evolution equation (IREE). The approach based on the reggeon Green function   
is discussed in Section 6. Some numerical estimates are given in Section 7.   
Section 8 contains our conclusions. Two     
appendixes contain some details of our calculations,     
supplementing sections 4 and 5.

\section{The quark box}

We begin with the lowest order diagrams for     
the scattering amplitude $T^{\gamma^*\gamma^*}$ of the elastic     
$\gamma^* \gamma^*$ scattering process. We restrict ourselves to the    
forward direction $t=0$, and for simplicity we first take the virtualities     
of all external photons to be equal.

The exact computation can be found in Ref.\cite{Budnev}; we restrict ourselves     
to the high energy behavior.     
The lowest order consists of the three fermion-loop    
diagrams (Fig. \ref{born}, a - c);    
at high energies only the planar box (a) and its     
$s \leftrightarrow u$ cross partner (b) contribute to the leading     
double-logarithmic behavior.       
    
 \FIGURE{   
\begin{tabular}{c c c }   
\epsfig{file=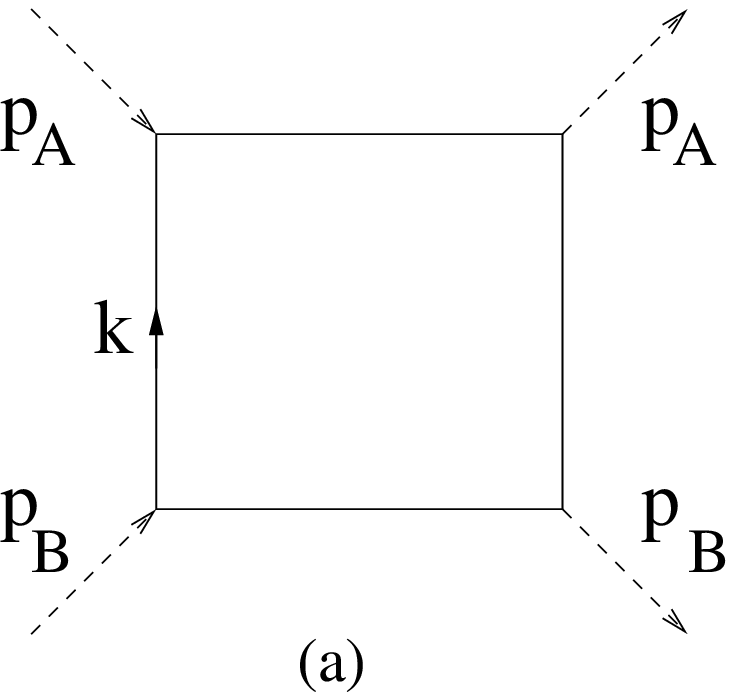,width=46mm}&   
\epsfig{file=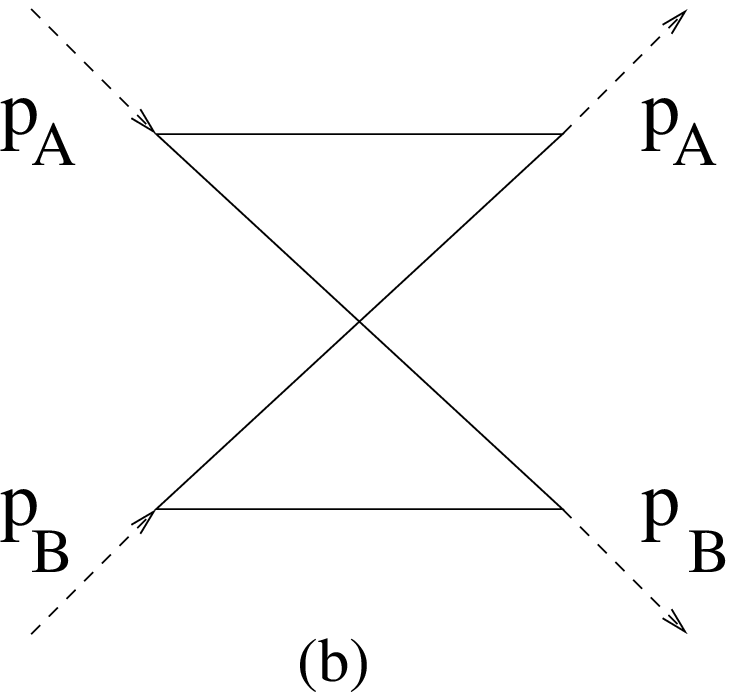,width=46mm}&   
\epsfig{file=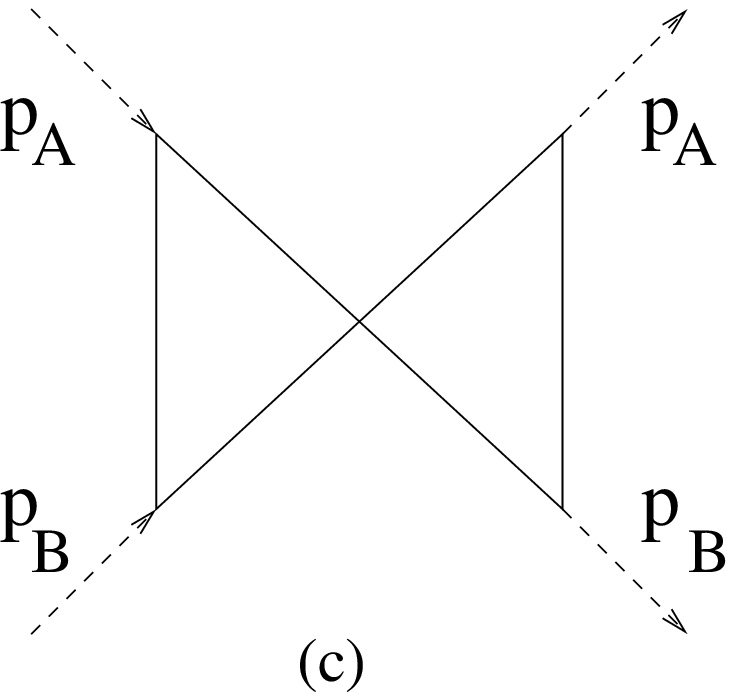,width=46mm} \\   
\end{tabular}   
  \caption{The Born level diagrams.}   
    \label{born}}

The calculations below are done using the Feynman gauge.    
Our method of extracting the double-logs will be close to the original     
paper ~\cite{GGL}. We introduce the notation (Fig. \ref{born})     
\begin{equation}    
p_A^2\,=\,-\,Q^2\,;\,\,\,\,\,\,\,\,p_B^2\,=\,-\,Q^2\,;\,\,\,\,\,\,\,\,\,\,    
(p_A\,+\,p_B)^2\,=\,s\,;\,\,\,\,\,\,\,\,\,\,x\,=\,Q^2/s\,.\,\,\,\,\,\,\,\,\,.    
\end{equation}    
In our Sudakov decomposition $k=\beta q - \alpha p + k_{\perp}$     
(with $k_{\perp}^2 = - \vec{k}^2$) the light cone vectors     
$p$, $q$ are defined through    
\begin{equation}    
p_A\,=\,p\,-\,x\,q\,;\,\,\,\,\,\,\,\,\,\,\,\,\,\,    
p_B\,=\,q\,-\,x\,p\,;\,\,\,\,\,\,\,\,\,\,\,\,\,    
p^2\,=\,q^2\,=\,0\,;\,\,\,\,\,\,\,\,\,\,\,\,\,    
2\,(p\,q)\,=\,s\,.    
\end{equation}    
Transverse polarization vectors are defined by    
$\epsilon_{\pm}^{\mu}\,=\,    
\frac{1}{\sqrt{2}}\, \left(0,1,\pm i,0\right)$, the longitudinal ones by    
$\epsilon_L(p_A)\,=\,\frac{1}{\sqrt{Q^2}}\,\left(p\,+\,x\, q\right)    
\,=\,\frac{1}{\sqrt{Q^2}}\, \left(p_A\,+\, 2\,x\, q \right)$,  ~   
$\epsilon_L(p_B)\,=\,\frac{1}{\sqrt{Q^2}}\,\left(    
p\,+\,x\, q\right) \,=\,    
\frac{1}{\sqrt{Q^2}}\,\left(p_B\,+\, 2\,x\, q\right)$.      
    
We first analyze the quantum number structure of the t-channel.     
We Fierz-transform the upper part of the trace expression:    
\beq    
\label{Fierz}    
\gamma^{\nu} (\ksla + \psla _A) \gamma^{\mu}\, =\,     
 \left( (k+p_A)^{\nu} g^{\mu \rho} \,+\, (k+p_A)^{\mu} g^{\nu \rho}    
                        \,-\,(k+p_A)^{\rho} g^{\mu \nu} \right) \gamma_{\rho}    
                  \,+\,i \epsilon^{\nu \sigma \mu \rho} (k+p_A)_{\sigma}    
                        \gamma_5 \gamma_{\rho}.    
\eeq    
As discussed in ~\cite{RS}, the vector current generates the even     
signature $A_2$ exchange, the pseudovector current the $A_1$ exchange.     
The absence of the scalar, pseudoscalar, and tensor currents seems to     
indicate that, in $\gamma^*\gamma^*$ scattering, there is no scalar or     
pseudoscalar ($\pi$-like) exchange; note, however, that the first gluon rung     
inside the quark loop might contain the axial anomaly, i.e. the coupling     
of the t-channel pion-like state to two gammas is nonzero.   
In our subsequent analysis we will restrict ourselves to the vector    
contribution. The pseudovector exchange as well as the pseudoscalar    
exchange decouple from the helicity     
conserving scattering, i.e. is of no interest for the total cross section.    
    
Let us now turn to the quark loop. In order to find a    
double-logarithmic contribution we need to find, in the     
trace expression in the numerator, terms proportional the leading power of     
$s$ and to $k_{\perp}^2$. For the numerator of the planar box diagram with     
transverse polarization we obtain:    
\begin{equation}\label{tr2}    
tr\,\left(\esla(A) \,\ksla \,\esla(B) \,   
(\ksla-\psla _B)\,\esla (B')\,\ksla \,\esla(A')    
\,(\psla _A+ \ksla)\right)  \,\approx \,     
p_A \,\cdot \,p_B \,\,   
tr \,\left(\esla(A) \,\ksla \,\esla(B) \,\esla(B')\, \ksla \,\esla(A') \,   
\right),    
\end{equation}      
where, on the rhs, we have made use of the fact that,    
in the high energy limit, the leading power of    
$s$ is due to the product $p_A \cdot p_B$. For the helicity conserving    
scattering we obtain:    
\begin{eqnarray}    
\label{tr3} & &   
tr\,\left(\esla(A)\, \ksla\, \esla(B) \,   
(\ksla-\psla _B)\,\esla (B')\,\ksla \,\esla(A')    
(\psla _A+ \ksla)\right) \,\approx \,2 \; s \; k_{\perp}^2 \,\times\nonumber\\    
& & \\    
& &    
\left[ \epsilon(A)\cdot \epsilon(A') \;\;  \epsilon(B)\cdot\epsilon(B')\;\;    
+ \;\;\epsilon(A)\cdot \epsilon(B') \;\; \epsilon(B)\cdot\epsilon(A')\;\;    
- \;\; \epsilon(A)\cdot \epsilon(B) \;\;\epsilon(A')\cdot\epsilon(B') \right].    
\nonumber   
\end{eqnarray}    
For the helicity conserving case (which we will need for the total cross    
section) the last two terms cancel. In the following, therefore, we shall    
restrict ourselves to the first term of the rhs of (\ref{tr3}).        
For later convenience we define the trace factor:    
\beq\label{TR}    
\tau_{TT}\,=\,4\,\,N_c\,\,\alpha_{em}^2\,\, F_{ns} \;\;     
\epsilon(A)\,\cdot \,\epsilon(A') \;\;     
\epsilon(B)\,\cdot\,\epsilon(B'),     
\eeq    
where    
\beq      
F_{ns}\,\,= \,\,   
\sum_{quarks} e_q^4\, \,-\, \,\frac{1}{N_f}\, (\sum_{quarks} e_q^2)^2    
\eeq    
denotes the projection on the flavor nonsinglet t-channel (for the flavor    
group $SU(N_f)$), and    
$e_q$ stands for the electric charge of the quark, measured in units of $e$.    
    
Diagram Fig. \ref{born}b    
will be obtained by simply substituting, at the end of our     
calculation, $s \to u$.     
The nonplanar diagram Fig. \ref{born}c has no contribution proportional to     
$s k_{\perp}^2$ and will be neglected.      
    
As to the integration over the momentum $k$,    
it will be instructive to first follow~\cite{GGL}: taking into account     
the $k_{\perp}^2$ factor from the numerator, we first do the     
transverse integration (formally by replacing one of the exchange     
propagators by a $\delta$ function),    
and are then left with the $\alpha$ and $\beta$-integrals     
of the type     
\beq\label{dl}    
\int \frac{d \alpha}{\alpha} \int \frac{d \beta}{\beta}    
\eeq    
which are restricted to lie in the region:    
\begin{equation}\label{limits}    
x \,<\, \alpha \,<\, 1,\;\; \,\,\,\,\,\,\,\,\,\,\,\,\,\,   
x\,<\, \beta \,<\, 1, \;\;\,\,\,\,\,\,\,\,\,\,\,\,\,   
 \mu_0^2 \,< \,s\,\alpha\, \beta\, =\, \vec{k}^2 \,<\, s       
\end{equation}    
where $\mu_0$ denotes the momentum scale which separates the infrared     
(nonperturbative) region from the hard region. In the following, however,     
we will use, as integration variables, $\beta$ and $\vec{k}^2$,    
and the limits     
have to be derived from (\ref{limits}).    
We have to distinguish between the two     
kinematic regions which we denote by `$+$' and `$-$':    
\beq\label{plusext}    
I^+:\;\; \mu_0^2 \,< \,\frac{Q^4}{s}\\    
\eeq    
    
\beq\label{minusext}    
I^-:\;\; \frac{Q^4}{s} \,<\, \mu_0^2.     
\eeq

In the first region, $I^+$, we have the following limits of integration    
\begin{equation}\label{result1plus}    
\int_{Q^2/s}^1 \frac{d \beta}{\beta} \int_{\beta Q^2}^{\beta s}    
\frac{d k^2}{k^2} \,\,   
= \,\,\ln ^2 \frac{s}{Q^2}.      
\end{equation}

In the second region, $I^-$, we have two separate contributions:    
\begin{equation}\label{result1minus}    
\int_{Q^2/s}^{\mu_0^2/Q^2} \frac{d \beta}{\beta}     
\int_{\mu_0^2}^{\beta s} \frac{d k^2}{k^2} \,+\,     
\int_{\mu_0^2/Q^2}^{1} \frac{d \beta}{\beta}     
\int_{\beta\,Q^2}^{\beta s} \frac{d k^2}{k^2}    
\,=\, \frac{1}{2} \left( \ln^2 \frac{s}{Q^2} \,-\,   
 \ln^2 \frac{Q^2}{\mu_0^2} \right)    
     \,+\, \ln \frac{s}{Q^2} \ln \frac{Q^2}{\mu_0^2}.    
\end{equation}

One easily verifies that at $\mu_0^2 \,=\, \frac{Q^4}{s}$ the two results     
coincide. All other photon polarizations, at high energies, are suppressed     
compared to the helicity conserving case of transverse polarization.    
For example, the case $TL \to TL$ goes as $const$, the case $LL \to LL$ as     
$1/s$. In the following we will restrict ourselves to the leading      
polarization, to $TT \to TT$.     
Combining (\ref{TR}) and (\ref{result1plus}), (\ref{result1minus})     
we write our result for the planar box diagram in the form     
\beq    
\label{box1}     
T_{box}^{\pm}\,=\, \,\tau_{TT} \,\left\{    
\begin{array}{ll}    
\ln ^2 \frac{s}{Q^2} & \;\;\;\,\,if\;\;\;\,\,\mu_0^2 \,\,<\,\, \frac{Q^4}{s}\\    
 \ln^2 \frac{s}{Q^2} - \frac{1}{2} \left( \ln \frac{s}{Q^2}     
- \ln \frac{Q^2}{\mu_0^2}\right)^2  &\,\, \;\;\; if \,\,\;\;\;   
\frac{Q^4}{s}\,\, <\,\, \mu_0^2    
\end{array} \right\}\,.     
\eeq

The nonplanar box in Fig. \ref{born}b    
is obtained by substituting $s \to u$, and in the     
sum of the two diagrams the obtained results stand for the even signature    
$A_2$ exchange.        
The total cross section for $\gamma^*\,\gamma^*$ (averaged over the incoming    
transverse helicities) follows from    
\beq    
\label{sigma}    
\sigma^{\gamma^*\,\gamma^*}\,=\,\frac{1}{s}\,Im\,T\,    
\simeq\,\frac{1}{s}\,\frac{\pi\,\D T}{\D \ln s}    
\eeq     
(with the last approximate equality being valid in the high energy approximation    
only) and has the form     
\beq    
\label{sigbox}    
\sigma^{\gamma^*\,\gamma^*}_{Born}\,    
=\,\tau_{TT} \,\pi\, \left\{     
\begin{array}{ll}    
2\;\ln \frac{s}{Q^2} & \,\,\,\,\,if\;\;\;\,\,\,\,\,\,\,\,\,\,   
\mu_0^2 \,\,<\,\, \frac{Q^4}{s}\\    
\ln \frac{s}{\mu_0^2}  & \,\,\,\,\,   
if \,\,\,\,\,\;\;\;\,\,\,\,\,\frac{Q^4}{s} \,\,<\,\, \mu_0^2    
\end{array}     
\right\}\,.    
\eeq    
    
It is not difficult to generalize our analysis to the case of unequal     
photon masses, $Q_1^2$ and $Q_2^2$. Instead of the scaling variable     
$x=Q^2/s$ we now have the two variables $x_i=Q_i^2/s$ ($i=1,2$). The     
integration region (\ref{limits}) is replace by     
\beq \label{limits12}        
x_2 <\, \alpha \,< \,1,\;\;\,\,\,\,\,\,\,\,\,\,\,\,\,\,   
 x_1\,< \,\beta \,< 1, \;\;\,\,\,\,\,\,\,\,\,\,  \,\,\,\,\,\,    
\mu_0^2 \,<\, s\,\alpha\, \beta \,=\,     
\vec{k}^2 \,<\, s\,.    
\eeq       
The two kinematic regions (\ref{plusext}), (\ref{minusext}) become    
\beq\label{plusext12}    
I^+:\;\; \mu_0^2 \,< \,\frac{Q_1^2\,Q_2^2}{s}\\    
\eeq    
\beq\label{minusext12}    
I^-:\;\; \frac{Q_1^2\,Q_2^2}{s} \,<\, \mu_0^2,     
\eeq    
and the result (\ref{box1}) for the planar box diagram takes the form:    
\beq    
\label{box1,12}     
T_{box}^{\pm}\,=\, \,\tau_{TT} \,\left\{    
\begin{array}{ll}    
\ln \frac{s}{Q_1^2} \,\ln \frac{s}{Q_2^2} & \,\,\,\,\,if\;\;\;\,\,\,\,\,   
\mu_0^2 \,<\,     
             \frac{Q_1^2 \,Q_2^2}{s}\\    
\ln \frac{s}{Q_1^2} \,\ln \frac{s}{Q_2^2}     
    \,-\,\frac{1}{2} \,\ln ^2 \frac{s \mu_0^2}{Q_1^2 \,Q_2^2}     
           
& \,\,\,\,\, if \,\,\,\,\,\;\;\;\frac{Q_1^2 Q_2^2}{s} \,<\, \mu_0^2    
\end{array} \right\}\,.      
\eeq    
    
\section{Linear equations for the ladder}

We now turn to higher order corrections to the quark loop diagrams.\\    
    
For the gluons we will use Feynman gauge, and we make use of the discussion    
given in ~\cite{BER2,GGL,KiLi,Ry,BER1,KoLi}. When selecting those    
Feynman diagrams      
which in the high energy limit contribute to the double logarithmic     
behavior, one has to distinguish between even and odd signature.     
For the case of quark-quark scattering it has been shown~\cite{KiLi}     
that the even signature amplitude is described by ladder diagrams,    
whereas the odd signature amplitude has also non-ladder graphs.         
In this and in the following section we derive and discuss a linear     
integral equation for the even signature amplitude; so we can restrict     
ourselves to QCD ladder diagrams. The odd signature case can be derived     
from the IREE (to be discussed in section 5).      
    
Let us consider the diagram with one gluon rung (Fig. \ref{rung}a).    
\FIGURE{   
\begin{tabular}{c c  }   
\epsfig{file=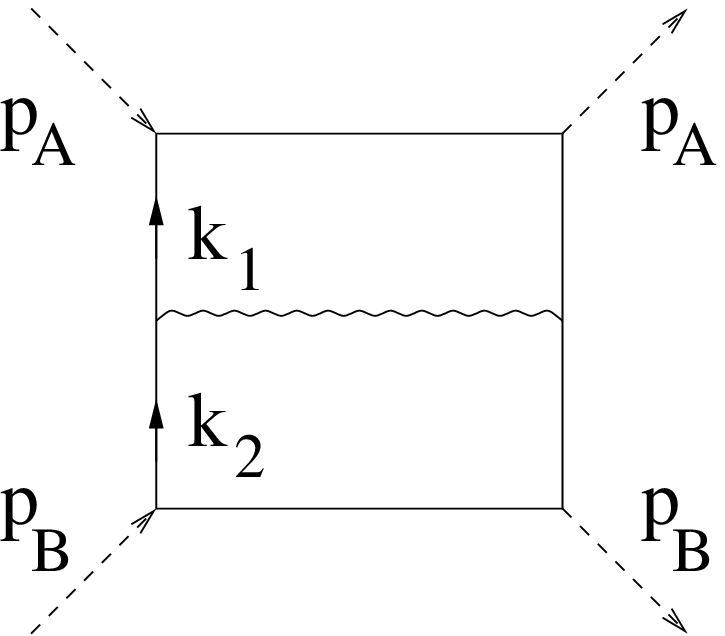,width=55mm} ~~~~~&   
~~~~~\epsfig{file=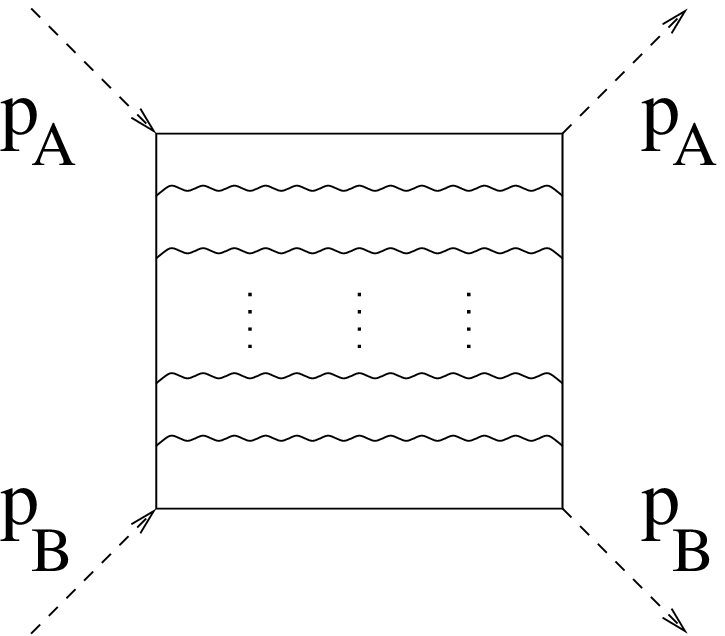,width=55mm} \\   
(a) & (b)\\   
\end{tabular}   
 \caption{ The first gluon rung (a). The ladder (b).}   
    \label{rung}}   
We begin with the     
trace and look for terms which result, for each of the two loop momenta     
$k_1$ and $k_2$, in a factor $\vec{k}_i^2$. Writing down the trace we    
find, for the lower cell, the product $...\gsla_{\mu}\; \ksla_2\; \esla(B)\;    
(\ksla\,-\,\psla_B)\;    
\esla(B')\; \ksla_2\; \gsla^{\mu}\;...$ which can be rewritten as    
$2 \,k_{2\;\perp}^2 \,\ksla_2\; \esla(B)\;    
(\ksla\,-\,\psla_B)\;    
\esla(B')\; \ksla_2 +...$ where the remainder (indicated by dots), for   
the helicity conserving case $\epsilon(B)=\epsilon(B')$, does not contribute    
to the double logarithmic approximation.    
The result for the trace therefore becomes      
\beq    
4\,s\, k_{1\;\perp}^2 \,k_{2\; \perp}^2 \,\epsilon(A)\,\cdot \,   
\epsilon(A')\;\;     
\epsilon(B) \,\cdot\, \epsilon(B').    
\eeq    
The $k^2$ factors will cancel against exchange propagators;    
together with color and $\pi$ factors we obtain the factor     
\beq    
\lambda\,\,=\,\,\frac{\alpha_s \,C_F}{2\, \pi}    
\eeq            
for the gluon rung.

Turning to the momentum integrations, we, once more, first follow the     
procedure outlined in ~\cite{GGL} and integrate over the transverse momenta.    
The remaining $\alpha$ and $\beta$ integrals are of the logarithmic type     
(\ref{tr3}), and the variables are ordered according to     
$x \,<\, \alpha_2\, <\, \alpha_1\, <\, 1$, ~   
$x \,<\, \beta_1\, <\, \beta_2 \,<\,1$, and     
their products are restricted by    
$\mu_0^2 \,<\, s\,\alpha_i\, \beta_i\, =\, \vec{k}_i^2\,<\, s$.    
In what follows, however, we choose the variables $\beta_i$, $k_i^2    
\,=\,\vec{k}_i^2$; the integrals to be done are     
\beq    
\int \frac{d \beta_1}{\beta_1} \,\,\int \frac{d k_1^2}{k_1^2}      
\,\,\int \frac{d \beta_2}{\beta_2} \,\,\int \frac{d k_2^2}{k_2^2}    
\eeq    
with the integration region    
\beq\label{limitrung}    
x\,=\,\frac{Q^2}{s} \,<\, \beta_1\, <\, \beta_2\, <\, 1,\;\;\,\,    
max(\beta_2 \,Q^2, \,\mu_0^2) \,<\, k_2^2\, <\, \beta_2\, s,    
\,\,\;\; max(\frac{k_2^2 \,\beta_1}{\beta_2},\,\mu_0^2) \,<\, k_1^2\, <\,   
 \beta_1\, s\,.     
\eeq    
As in the one loop case, we shall see that the two possibilities    
(\ref{plusext}) and (\ref{minusext}) will have to be distinguished.     
We start with      
the integration of the upper cell, keeping the variables of the lower cell,     
$\beta_2$, $k_2^2$, fixed.    
The restrictions (\ref{limitrung}) imply two distinct regions    
(Fig. \ref{int1}):    
\beq\label{plus2}    
I_2^+:\;\;\; \mu_0^2 \,\,<\,\, \frac{Q^2 \,k_2^2}{s \,\beta_2} \\     
\eeq    
\beq\label{minus2}    
I_2^-:\;\;\; \frac{Q^2\, k_2^2}{s \,\beta_2}\,\, <\,\, \mu_0^2\,.    
\eeq    
    
\FIGURE{   
\begin{tabular}{c c }   
\epsfig{file=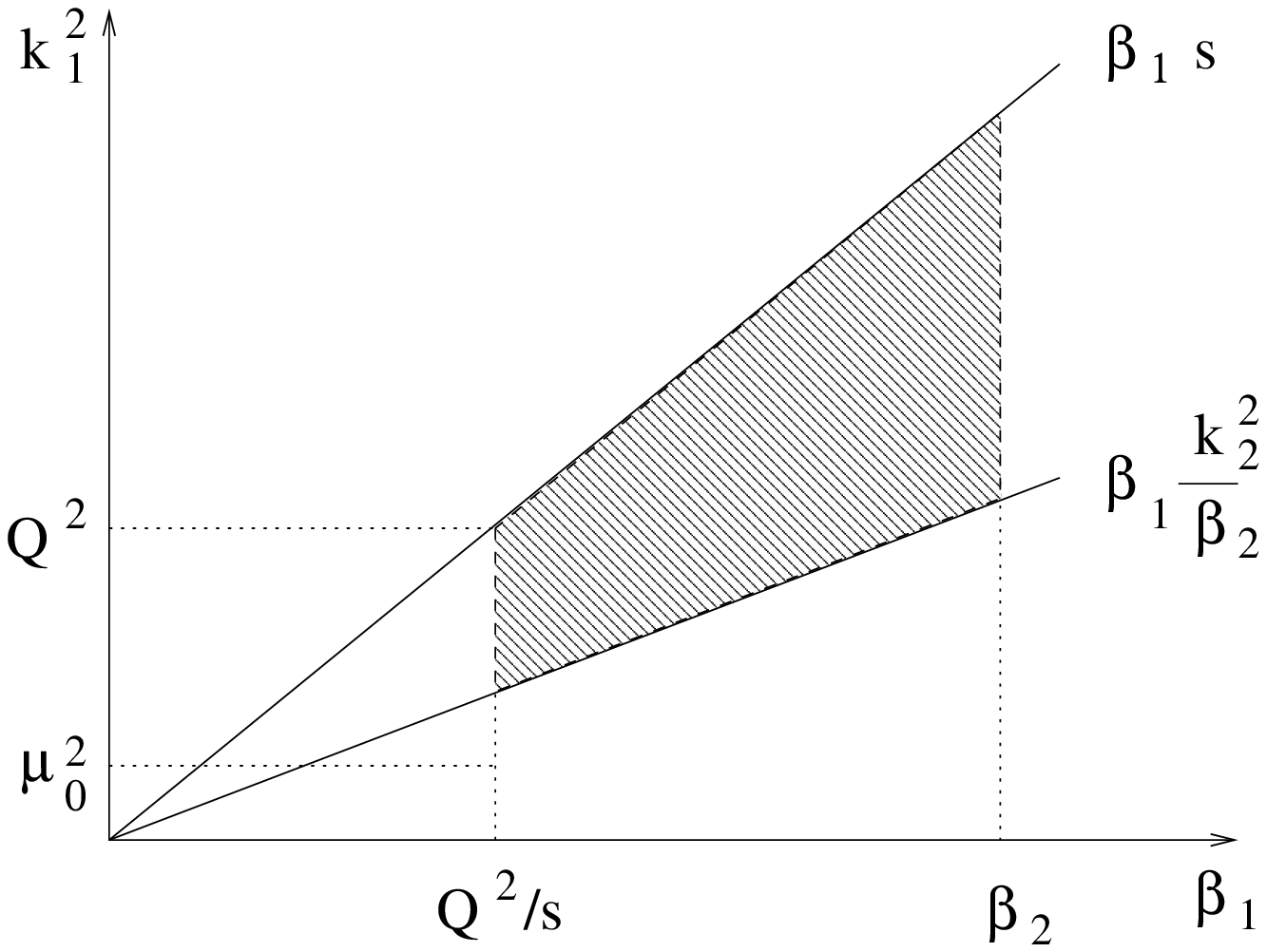,width=70mm}&   
\epsfig{file=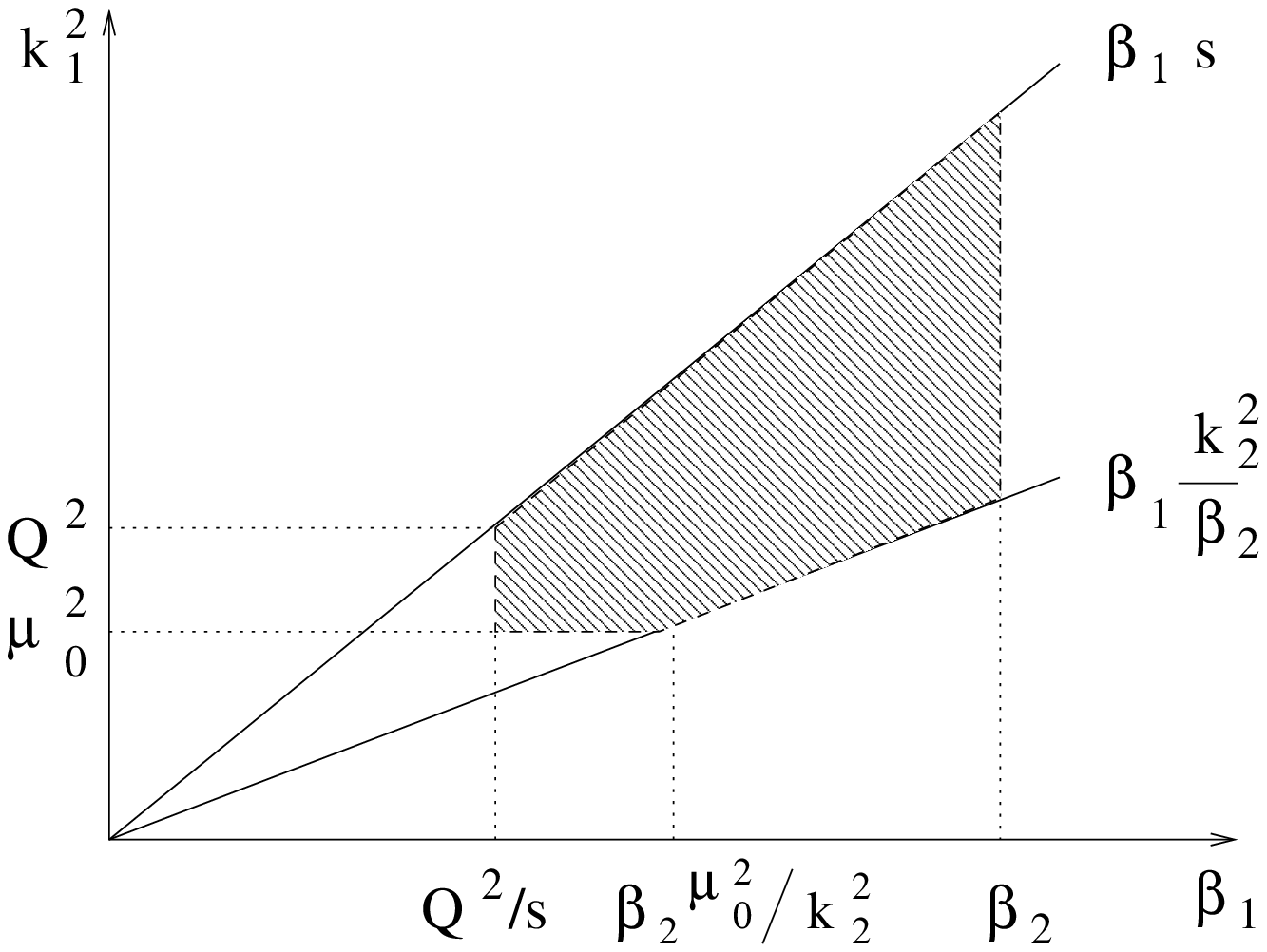,width=70mm}\\   
(a) ~ $\mu_0^2\,<\,\frac{k_2^2\,Q^2}{\beta_2\,s}$ &   
(b) ~ $\mu_0^2\,>\,\frac{k_2^2\,Q^2}{\beta_2\,s}$ \\   
\end{tabular}   
  \caption{ The integration domains for one rung: upper loop.}   
    \label{int1}}   
   
In the first case we have     
\beq\label{b1}   
\int_{Q^2/s}^{\beta_2} \,\frac{d \beta_1}{\beta_1} \,   
\int_{\beta_1\, k_2^2 / \beta_2}^{s \,\beta_1} \,\frac{d k_1^2}{k_1^2}    
\,\,=\,\, \ln \frac{s\,\beta_2}{Q^2}\, \ln \frac{s\, \beta_2}{k_2^2},     
\eeq    
whereas the second one leads to a sum of two contributions:    
\begin{eqnarray}\label{b2}    
 \int_{Q^2/s}^{\beta_2 \,\mu_0^2/k_2^2} \,\frac{d \beta_1}{\beta_1}     
\,\int_{\mu_0^2}^{\beta_1\, s}\, \frac{d k_1^2}{k_1^2}\,\, +\,\,     
\int_{\beta_2 \,\mu_0^2/k_2^2}^{\beta_2} \,\frac{d \beta_1}{\beta_1}    
\,\int_{\beta_1\, k_2^2/ \beta_2}^{s\,\beta_1}    
\frac{d k_1^2}{k_1^2}\nonumber\\    
=\,\, \frac{1}{2} \,   
\left(\ln ^2 \frac{\beta_2\, s}{k_2^2} \,\,-\,\, \ln ^2 \frac{Q^2}{\mu_0^2}    
\right) \,\,+\,\, \ln \frac{k_2^2}{\mu_0^2}\, \ln \frac{\beta_2 \,s}{k_2^2}\,.    
\end{eqnarray}

For the remaining integrations over $\beta_2$ and $k_2^2$ (Fig. \ref{int2})    
we not only     
have to observe the distinction (\ref{plus2}), (\ref{minus2}),     
but also to return to     
the two cases defined in (\ref{plusext}), (\ref{minusext}).    
 \FIGURE{   
\begin{tabular}{c c }   
\epsfig{file=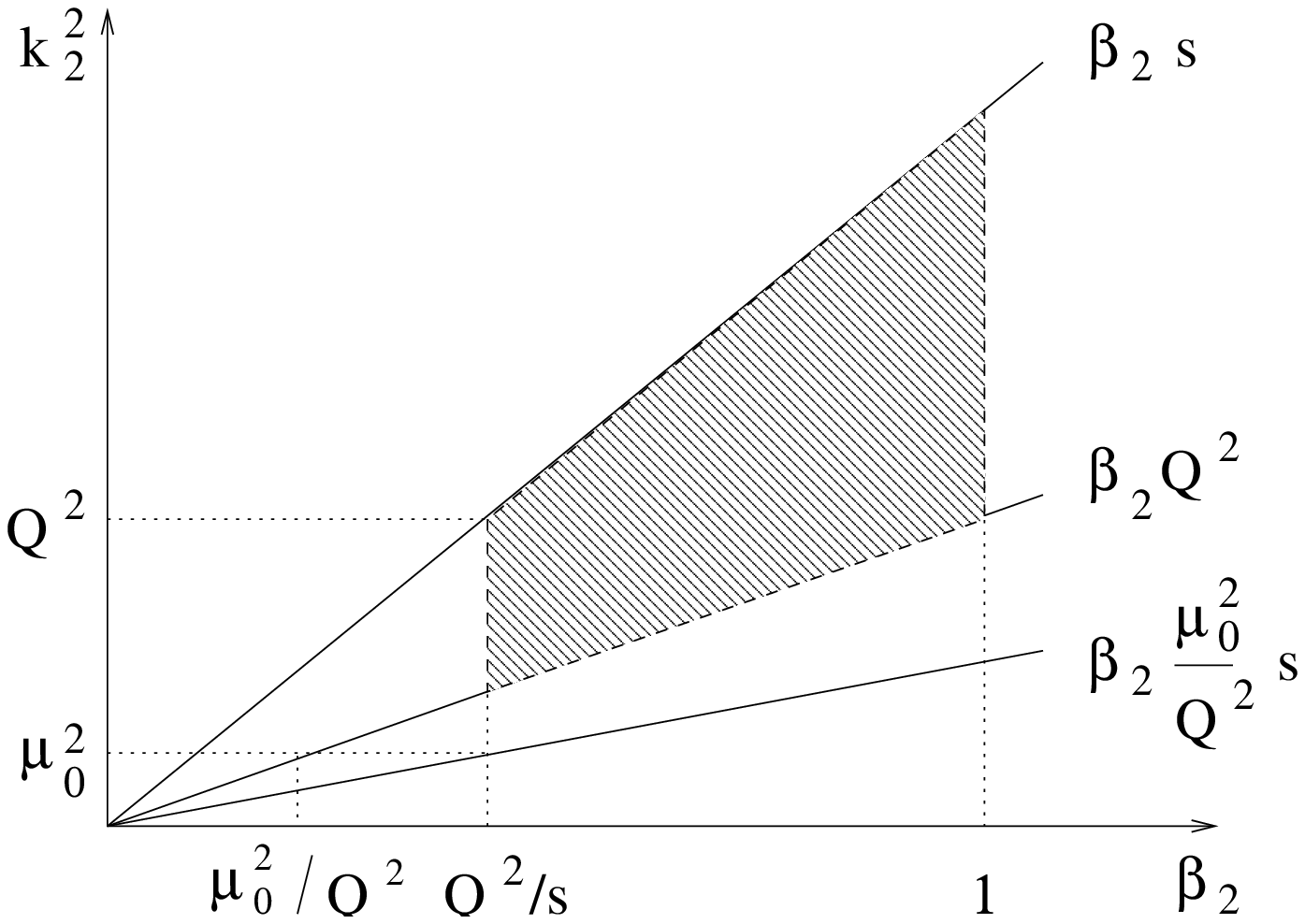,width=70mm} &   
\epsfig{file=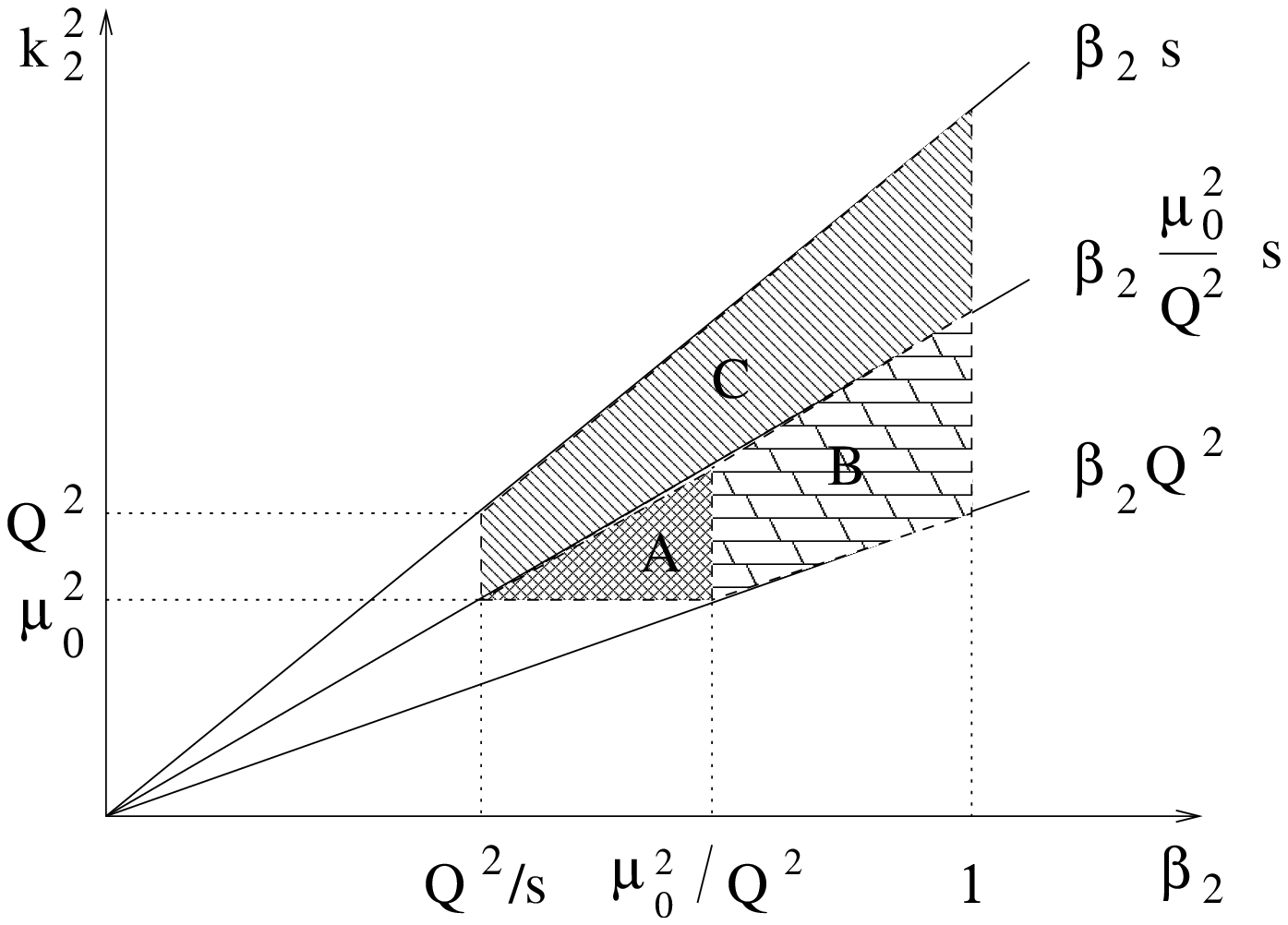,width=70mm} \\   
(a) $\mu_0^2\,<\,\frac{Q^4}{s}$ &   
(b) $\mu_0^2\,>\,\frac{Q^4}{s}$ \\   
\end{tabular}   
  \caption{ The integration domains for one rung: lower loop.}   
    \label{int2}}   
In the first case (i) we find (Fig. \ref{int2}a)     
that the line $k_2^2 \,=\, \beta_2\, Q^2$ always lies above the line     
$\beta_2 \,\mu_0^2\, s/ Q^2$ and    
$max(\beta_2\, Q^2,\, \mu_0^2)\,=\, \beta_2 \,Q^2$.     
Therefore in the whole integration region (shaded area in Fig. \ref{int2}a)     
we have $k_2^2/s\, >\, \beta_2\, \mu_0^2 /Q^2$, and the first condition    
  (\ref{plus2})   
is always fulfilled. This leads to      
\beq\label{result2plus}     
\int_{Q^2/s}^{1}\, \frac{d \beta_2}{\beta_2}\,    
\int_{\beta_2 \,Q^2}^{\beta_2\, s} \,   
\frac{d k_2^2}{k_2^2}\, \ln \frac{\beta_2 \,s}{Q^2} \,   
\ln \frac{\beta_2\, s}{k_2^2}\, =\,    
\frac{1}{4}\, \ln ^4 \,\frac{s}{Q^2}\,.     
\eeq

For the second case (ii) in (\ref{minus2})     
the situation is slightly more complicated       
(Fig. \ref{int2}b):    
the region of integration (shaded area) consists of three pieces    
(denoted by A, B, and C). In C the first condition (\ref{plus2}) holds,     
whereas     
in B and C the second one (\ref{minus2}) is fulfilled.     
For region A the result is:        
\begin{eqnarray}\label{result2minus}    
 \int_{Q^2/s}^{\mu_0^2/Q^2}\, \frac{d \beta_2}{\beta_2} \,    
\int_{\mu_0^2}^{\beta_2 \,s\, \mu_0^2/Q^2} \,\frac{d k_2^2}{k_2^2}    
\,\left(     
\frac{1}{2} \,\left( \ln ^2 \frac{\beta_2\, s}{k_2^2}\, -\,     
\ln ^2 \frac{Q^2}{\mu_0^2} \right) \,+\,     
\ln \frac{k_2^2}{\mu_0^2}\, \ln \frac{\beta_2 \,s}{k_2^2} \right)     
\nonumber\\    
=\,\,\frac{1}{12} \,\ln ^4 \frac{s}{Q^2}  \,   
-\, \frac{1}{2}\, \ln ^2 \frac{s}{Q^2} \,\ln ^2 \frac{Q^2}{\mu_0^2}    
\,+\, \frac{2}{3} \,\ln \frac{s}{Q^2}\, \ln ^3 \frac{Q^2}{\mu_0^2}    
\,-\,\frac{1}{4}\, \ln ^4 \frac{Q^2}{\mu_0^2}\,.     
\end{eqnarray}    
For region B we obtain:    
\begin{eqnarray}    
 \int_{\mu_0^2/Q^2}^{1} \,\frac{d \beta_2}{\beta_2}\,     
\int_{\beta_2\, Q^2}^{\beta_2\, s\, \mu_0^2/Q^2}\, \frac{d k_2^2}{k_2^2} \,   
\left( \frac{1}{2}\,    
\left( \ln ^2 \frac{\beta_2\, s}{k_2^2}\, -\, \ln ^2 \frac{Q^2}{\mu_0^2}    
\right) \,+\, \ln \frac{k_2^2}{\mu_0^2} \,   
\ln \frac{\beta_2 \,s}{k_2^2}\right)\nonumber \\    
\,\,=\,\, \frac{1}{3} \,\ln ^3 \frac{s}{Q^2}\, \ln \frac{Q^2}{\mu_0^2}    
 \, +\, \frac{1}{4}\, \ln ^2 \frac{s}{Q^2} \,\ln ^2 \frac{Q^2}{\mu_0^2}    
\,-\, \ln  \frac{s}{Q^2}\, \ln ^3 \frac{Q^2}{\mu_0^2}\,    
+\, \frac{5}{12}\, \ln ^4 \frac{Q^2}{\mu_0^2}\,.    
\end{eqnarray}    
Finally, region C:    
\begin{eqnarray}    
 \int_{Q^2/s}^{1} \,\frac{d \beta_2}{\beta_2}  \,   
\int_{\beta_2\, s\, \mu_0^2/Q^2}^{\beta_2 \,s}\, \frac{d k_2^2}{k_2^2}    
\,\ln \frac{s\,\beta_2}{Q^2}\, \ln \frac{s \,\beta_2}{k_2^2}    
\,=\,\frac{1}{4} \,\ln ^2 \frac{s}{Q^2}\, \ln ^2 \frac{Q^2}{\mu_0^2}\,.    
\end{eqnarray}    
Adding up these three contributions we find    
\beq \label{rung2}   
\frac{1}{12} \,\left( \ln ^4 \,\frac{s}{Q^2} \,+\, 2\,    
\ln ^4 \frac{Q^2}{\mu_0^2} \right)    
\,+\,\frac{1}{3} \left( \ln ^3 \frac{s}{Q^2}\, \ln \frac{Q^2}{\mu_0^2} \,-\,    
\ln \frac{s}{Q^2} \,\ln ^3 \frac{Q^2}{\mu_0^2} \right) \,.    
\eeq     
At the point $\mu_0^2 \,=\,\frac{Q^4}{s}$    
both results (\ref{result2plus}) and (\ref{rung2}) coincide.

After the analysis of the one rung ladder diagram it is not difficult to     
see the general pattern and to derive an integral equation. The general     
diagram with $n$ rungs is illustrated in Fig. \ref{rung}b.    
>From the trace expression     
in the numerator we obtain, for each rung, a factor    
$k_{i \perp}^2 \,\lambda$,     
where the momentum factor cancels against one of the exchange propagators     
of the cell labeled by ``i''.    
Following the strategy adopted for the one loop     
diagram and for the two loop case, we arrive at the $\alpha$ and $\beta$     
integrals of the logarithmic type (\ref{dl}), with the ordering     
conditions     
\beq    
x \,<\, \alpha_n \,<\, \alpha_{n-1} \, ...\, \alpha_1 \,<\, 1,    
\;\;\;\,\,\,\,\,\,\,\,\,\,\,\,    
x\, <\, \beta_1\, <\, \beta_2\, <\, ...\, <\, \beta_{n-1}\,    
< \,\beta_n \,< \,1    
\eeq    
and the additional constraints    
\beq    
\mu_0^2 \,<\, s\, \alpha_i \,\beta_i\, =\,  \vec{k}_i^2\, <\, s \,.   
\eeq         
Translating to the variables $\beta_i$ and    
$k_i^2\,=\,\vec{k}_i^2$, the limits     
of integration of the $k_i^2$ become, in analogy with (\ref{limitrung}):    
\begin{eqnarray} \label{ord}   
max(\beta_n \,Q^2, \,\mu_0^2) \,<\, k_n^2 \,<\, \beta_n\, s,\;\,\,\,\,\,\,\,    
max(\frac{\beta_{n-1} \,k_n^2}{\beta_n},\, \mu_0^2) \,< \,k_{n-1}^2\, <\,     
\beta_{n-1}\, s\;...\,,\, \nonumber \\    
max(\frac{\beta_1 \,k_2^2}{\beta_2}, \,\mu_0^2)\, <\, k_1^2\, <\,    
\beta_1 \,s\,.    
\end{eqnarray}    
Starting from cell 1 at the top, we now work our way down along the ladder.    
We first do the integrals of cell 1, keeping fixed the variables of all the     
other cells. In doing so, we are led to distinguish the two cases     
(\ref{plus2}) and (\ref{minus2}); the results are given in (\ref{b1}) and     
(\ref{b2}), which we denote by $A_1^+ (\beta_2,\,k_2^2)$ and     
$A_1^- (\beta_2,\,k_2^2)$, resp. The following integration of cell 2     
is done as described before: the only difference is that (\ref{plusext}) and     
(\ref{minusext}) are replaced by     
\begin{eqnarray}     
I_3^+:\;\;\; \mu_0^2 < \frac{Q^2 k_3^2}{s \beta_3}\,;\\     
I_3^-:\;\;\; \frac{Q^2 k_3^2}{s \beta_3} < \mu_0^2\,.         
\end{eqnarray}     
As before, the `$+$' region receives contributions only from the `$+$'   
region (Fig.\ref{int2}a), whereas     
the `$-$' region has contributions from both region A and B     
and from region C (Fig.\ref{int2}b). Denoting the results by     
$A_2^+ (\beta_3,\,k_3^2)$ and     
$A_2^- (\beta_3,\,k_3^2)$, we have, symbolically,    
\beq    
A_2^+ \,\,=\,\, K^{++}\,\, \otimes \,\,A_1^+    
\eeq    
and     
\beq    
A_2^- \,\,=\,\, K^{-+}\,\, \otimes \,\,A_1^+ \,\,+\,\, K^{--} \,\,   
\otimes\,\, A_1^- \,.   
\eeq     
Note, in particular, that the evolution of $A^+$ decouples from $A^-$.    
The same pattern now repeats itself for the integration of cell 3 etc,    
and we can write down a recursion relation. Instead, we define     
$A^{\pm}(\beta, \,k^2)\, = \,\sum_n \,A_n^{\pm}$    
(where $A_0^{\pm}\,=\,1$) which satisfy     
the coupled integral equations (Fig. \ref{BSfig}):    
\begin{equation}\label{BS+}    
A^+(\beta, \,k^2) \,=\,    
1\, + \,\lambda\,\int_{Q^2/s}^{\beta} \frac{d \beta'}{\beta'}    
      \int_{\beta' \,k^2/\beta}^{\beta'\, s} \,\frac{d k^{'2}}{k^{'2}}    
               \, A^+(\beta', \,k^{'2})    
\end{equation}       
and     
\begin{eqnarray}\label{BS-}    
&&A^- (\beta,\, k^2)\,\,=\,\,   1\, +\,\lambda\,    
                \int_{Q^2/s}^{\beta} \frac{d \beta'}{\beta'}    
             \,   \int_{\beta'\, \mu_0^2/Q^2}^{\beta' s}\,   
 \frac{d k^{'2}}{k^{'2}}\,    
                A^+(\beta',\, k^{'2})\,\,+ \nonumber \\ &  \\    
               &&\lambda    
        \int_{Q^2/s}^{\beta \,\mu_0^2/ k^2} \frac{d \beta'}{\beta'}    
                \int_{\mu_0^2}^{\beta'\, \mu_0^2 s/Q^2}   
 \frac{d k^{'2}}{k^{'2}} \,   
                A^- (\beta',\, k^{'2})    
                \,+ \,\lambda\,    
\int_{\beta \,\mu_0^2/k^2}^{\beta}\, \frac{d \beta'}{\beta'}    
                \int_{\beta' \,k^2/\beta}^{s \,\beta' \,\mu_0^2/Q^2}     
                \frac{d k^{'2}}{k^{'2}}\,    
                A^-(\beta', \,k^{'2}) . \nonumber    
\end{eqnarray}    
 \FIGURE{\epsfig{file=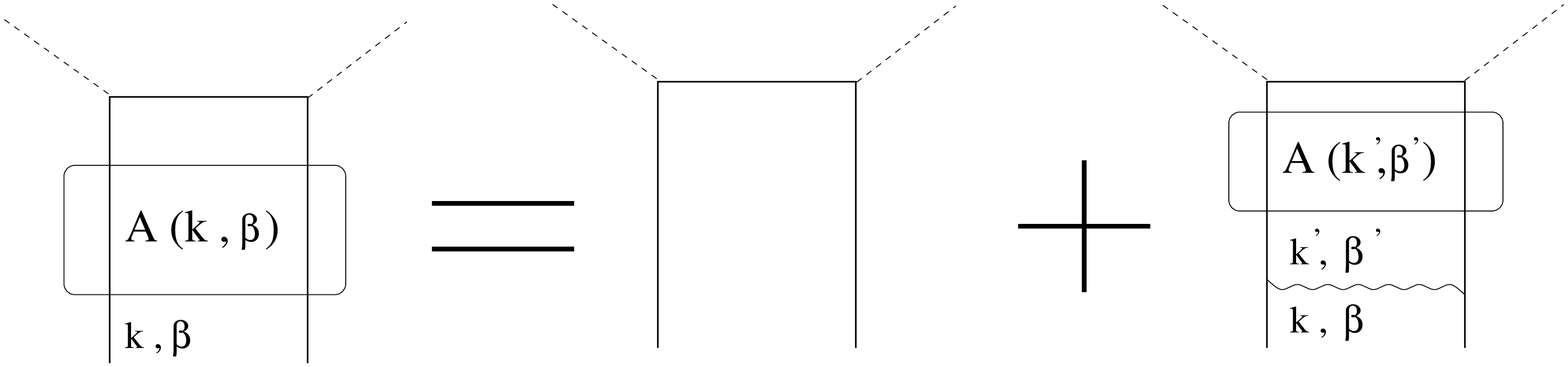,width=140mm}   
 \caption{ The Bethe - Salpeter equation.}   
    \label{BSfig}}   
    
The first equation applies to the region $I_A^+$, the second one to $I_A^-$.    
They are defined by    
\beq\label{plus}    
I_A^+:\;\;\; \mu_0^2 \,\,< \,\,\frac{Q^2\, k^2}{\beta \,s}    
\eeq    
\beq\label{minus}    
I_A^-:\;\;\; \frac{Q^2 \,k^2}{\beta \,s}\,\, <\,\, \mu_0^2\,.    
\eeq     
    
Finally, the amplitudes $T^{\pm}$ for the    
photon-photon scattering are obtained from $A$ by subtracting the     
terms $A_0^{\pm}\,=\,1$ and by putting $k^2\,=\,Q^2$ and    
$\beta\,=\,1$:    
    
\beq\label{amp} T^{\pm}(Q^2,\,s)\,=\,    
\left[A^{\pm}(1,\,Q^2)\,-\,1\right]\,\frac{\tau_{TT}}{\lambda}\,.    
 \eeq    
Note that in this limit the regions $I_A^+$, $I_A^-$ coincide with     
$I^+$ and $I^-$ in (\ref{plusext}) and (\ref{minusext}).        
The full scattering amplitude is obtained by adding the twisted    
(with respect to $s \leftrightarrow u$ crossing) fermion loop.      
    
Before we finish this section we again briefly present the     
generalization to the case of unequal photon masses. With the main differences    
being the lower limits of the $\alpha$ and $\beta$ variables      
(analogous to (\ref{limits12})), the analysis for the one gluon rung and for    
the derivation of the integral equations can be done in exactly the same     
way as for the equal mass case; the final results for the integral equations     
are obtained from (\ref{BS+}) and (\ref{BS-}) by simply substituting     
$Q^2 \to Q_1^2$:      
\begin{equation}\label{BS+12}    
A^+(\beta, k^2) \,=\,    
1 \,+ \,\lambda\,\int_{Q_1^2/s}^{\beta} \frac{d \beta'}{\beta'} \,   
                \int_{\beta' \,k^2/\beta}^{\beta'\, s}\,   
 \frac{d k^{'2}}{k^{'2}} \,   
                A^+(\beta',\, k^{'2})    
\end{equation}       
and     
\begin{eqnarray}\label{BS-12}    
& &A^- (\beta, \,k^2) =   1\, +\,\lambda\,    
                \int_{Q_1^2/s}^{\beta} \frac{d \beta'}{\beta'}    
                \int_{\beta'\, \mu_0^2/Q_1^2}^{\beta'\, s}\,   
             \frac{d k^{'2}}{k^{'2}}    
                A^+(\beta',\, k^{'2}) \,+\nonumber \\ & & \\    
                &\lambda&    
        \int_{Q_1^2/s}^{\beta \,\mu_0^2/ k^2} \frac{d \beta'}{\beta'}   
                \int_{\mu_0^2}^{\beta'\, \mu_0^2\, s/Q_1^2}   
          \frac{d k^{'2}}{k^{'2}}\,    
                A^- (\beta',\, k^{'2})\,    
                + \,\lambda   
\int_{\beta\, \mu_0^2/k^2}^{\beta} \frac{d \beta'}{\beta'}    
                \int_{\beta' \,k^2/\beta\,s}^{s \,\beta'\, \mu_0^2/Q_1^2}     
                \frac{d k^{'2}}{k^{'2}}\,    
                A^-(\beta', \,k^{'2}). \nonumber    
\end{eqnarray}    
The same replacement applies to the kinematic regions (\ref{plus}),     
(\ref{minus}). At the end we put $k^2\,=\,Q_2^2$ and $\beta\,=\,1$:      
\beq\label{amp12}     
T^{\pm}(Q_1^2,\,Q_2^2,\,s)\,\,=\,\,    
\left[A^{\pm}(1,Q^2_2)\,-\,1\right]\,\frac{\tau_{TT}}{\lambda}.    
 \eeq    
For the remainder of our paper we will stay with this general case of unequal     
photon masses.      
    
\section{Solution of the linear equations}

The structure of the two equations, (\ref{BS+12}) and (\ref{BS-12})    
defines our strategy: we first solve the equation for $A^+$,    
Eq. (\ref{BS+12}),     
and we then use the solution as an inhomogeneous term in the equation for     
$A^-$, Eq. (\ref{BS-12}).    
    
\subsection{Solution in the region $I_A^+$}    
In the region $I_A^+$ where Eq. (\ref{BS+}) holds we define the new variables:    
$$\xi\,=\,\ln(\beta\, s/k^2)\,;\,\,\,\,\,\,\,\,\,\,\,\,\,\,\,   
\,\,\,\,\,\,\,\,\,\eta\,=\,\ln    
(\beta\,s/Q_1^2)\,.$$ In these new variables the equation (\ref{BS+})    
can be rewritten     
    
\beq\label{BS+1}    
A^+(\xi,\,\eta)\,=\,1\,+\,\lambda\,\int^{\eta}_{0}\, d\bar \eta\,   
\int^\xi_{0}\,    
d\bar\xi\,A^+(\bar\xi,\,\bar\eta)\,.    
\eeq

When differentiated  twice Eq. (\ref{BS+1}) reduces to    
\beq\label{BS+2} \frac{d^2 A^+}{d\xi\,d\eta}\,\,=\,\,\lambda\,A^+ \,.\eeq

A solution to Eq. (\ref{BS+2}) can be found in the following form    
\beq\label{sol+} A^+(\xi,\,\eta)\,=\,    
\int_{0^+}\frac{dz}{2\,\pi\,i}\,\phi_+(z)\,e^{\,\eta\,z\,+\,\lambda\,\xi/z}\,,    
\eeq with the integration path closed around the point $z\,=\,0$.    
    
The function $\phi_+(z)$ is uniquely determined when (\ref{sol+})    
is substituted into Eq. (\ref{BS+1}). The result is     
\beq\label{fi+}    
\phi_+(z)\,=\,\frac{1}{z},    
\eeq    
and    
\beq\label{sol+1}A^+(\xi,\,\eta)\,=\,I_0 \left    
(\sqrt{4\,\lambda\,\xi\,\eta}\right )\,.     
\eeq

\subsection{Solution in the region $I_A^-$}

In the region $I_A^-$  we have the integral equation (\ref{BS-})     
which couples the functions $A^+$ and $A^-$.

Again we define a new variable:

$$\xi^\prime\,=\,\xi\,-\,L_0\,;\,\,\,\,\,\,\,\,\,\,\,\,\,\,\,\,\,\,\,\,\,   
\,\,\,\,\,\,\,\,   
L_0\,=\,\ln (Q^2_1/\mu_0^2)\,.$$    
In the variables ($\xi^\prime ,\,\eta$) the equation (\ref{BS-})    
can be rewritten    
\begin{eqnarray}\label{BS-1}    
A^-(\xi^\prime,\,\eta)\,=\,1&+&\lambda\,\int^{\eta    
}_{0}\,d\bar\eta\,\int_{-L_0}^{0}\, d\bar \xi    
\,A^+(\bar\xi^\prime,\,\bar\eta) \nonumber \\   
&+&\lambda\,\int^{\xi^\prime    
}_{0}\,d\bar\eta\,\int^{\bar\eta}_{0}\, d\bar \xi^\prime    
\,A^-(\bar\xi^\prime,\,\bar\eta)\,+\,    
\lambda\,\int_{\xi^\prime    
}^{\eta}\,d\bar\eta\,\int^{\xi^\prime }_{0}\, d\bar \xi^\prime    
\,A^-(\bar\xi^\prime,\,\bar\eta)\, . \end{eqnarray}    
Note that, at $\xi'=0$, $A^-(0,\eta)$ coincides with $A^+(L_0,\eta)$:    
$0\,=\,\xi'\,=\,\xi \,- \,L_0 \,= \,   
\ln \frac{\beta \,s\,\mu_0^2}{k^2 \,Q_1^2}$ denotes the     
point where the two regions $I_A^+$ and $I_A^-$ touch each other.

After double differentiation (\ref{BS-1}) reduces to    
\beq\label{BS-2}    
\frac{d^2A^-}{d\xi^\prime\,d\eta}\,=\lambda\,A^- \,.\eeq    
    
For the solution we use the same ansatz as before:     
\beq\label{sol-} A^-(\xi^\prime,\,\eta)\,=\,    
\int_{0^+}\frac{dz}{2\,\pi\,i}\,\phi_-(z)\,e^{\,\eta\,z\,+\,\lambda\,\xi^\prime    
/z}\,, \eeq with the integration path closed around $z=0$.

The function $\phi_-(z)$ is determined when (\ref{sol-}) is    
substituted to Eq. (\ref{BS-1}). Note that  the boundary condition    
$A^-(0,\,\eta)\,=\,A^+(L_0, \,\eta)$ is respected by  Eq.    
(\ref{BS-1}). The result (see Appendix A) is     
\beq\label{fi-}    
\phi_-(z)\,=\,\frac{1}{z}\,e^{\lambda\,L_0/z}\,-\,\frac{z}{\lambda}\,e^{z\,L_0}\eeq    
and     
\beq\label{sol-1}A^-(\xi^\prime,\,\eta)\,=\,I_0\left    
(\sqrt{4\,\lambda\,(\xi^\prime\,+\,L_0)\,\eta}\right )    
\,+\,\frac{\xi^\prime}{\eta+L_0}\,I_2\left    
(\sqrt{4\,\lambda\,\xi^\prime\,(\eta\,+\,L_0)}\right ).    
\eeq

\subsection{Analyzing the solutions}    
    
Define    
$$\omega_0\,=\,\sqrt{4\,\lambda}\,;\,\,\,\,\,\,\,\,\,\,\,\,\,\,\,\,\,\,\,\,\,    
\,\,\,\,\,\,\,\,\tilde Q^2=\sqrt{Q_1^2\,Q_2^2}\,.$$    
    
The amplitude for $\gamma^*\,\gamma^*$ scattering is obtained using    
Eq. (\ref{amp})     
\beq\label{Tamp+} T^+\,=\,T(Q_1^2,\,Q_2^2,\,s)\,=\,    
\frac{4\;\tau_{TT}}    
{\omega_0^2}\,  
\left[I_0\left(\omega_0\,\sqrt{\ln\frac{s}{    
Q_1^2}\,\ln\frac{s}{Q_2^2}}\,\right)\,-\,1\right]\,    
\,\,\,\,\,\,\,\,\,\,if\,\,\,\, \mu_0^2\,<\,\tilde Q^4/s\eeq    
and    
\begin{eqnarray}\label{Tamp-} T^-&=&T(Q_1^2,\,Q_2^2,\,s)\,=\,   
\frac{4\;\tau_{TT}}{\omega_0^2} \,\times \nonumber \\    
& &\left [    
I_0\left(\omega_0\,\sqrt{\ln\frac{s}{    
Q_1^2}\,\ln\frac{s}{Q_2^2}}\,\right)\,-\;1 - \;\;\frac{\ln\frac{s\,\mu_0^2}{\tilde    
Q^4}}{\ln\frac{s}{\mu_0^2}}\,   
I_2\left(\omega_0\,\sqrt{\ln\frac{s\,\mu_0^2}{\tilde    
Q^4}\ln    
\frac{s}{\mu_0^2}}\right)\,\right]\,    
\nonumber \\ & &\,\,\,\,\,\,\,\,\,\,\,\,\,\,\,\,\,\,\,   
\,\,\,\,\,\,\,\,\,\,\,\,\,\,\,\,\,\,\,\,\,\,\,\,\,\,\,\,\,\,\,\,\,\,\,\,   
\,\,\,\,\,\,\,\,\,\,\,\,\,\,\,\,\,\,\,\,\,\,\,\,\,\,\,\,\,\,\,\,\,\,\,\,   
\,\,\,\,\,\,\,\,\,\,\,\,\,\,\,\,\,\,\,\,\,\,\,\,\,\,\,\,\,\,\,\,\,\,\,\,   
if\,\,\,\, \mu_0^2\,>\,\tilde    
Q^4/s\,.  
\end{eqnarray}

It is important to note that, although in the course of our derivation    
we seem to have lost the symmetry in $Q_1^2$ and $Q_2^2$, the    
final result of the amplitude is fully symmetric again. In particular, the    
second term depends upon $\tilde Q^4/s$ only.     
The amplitude $T^-$ reduces to $T^+$ when $\tilde Q^4/s = \mu_0^2$, i.e.     
when the dynamical infrared cutoff of the perturbative calculation     
reaches $\mu_0^2$, the limit of the nonperturbative infrared region.    
    
Let us consider, in some more detail, the $s\rightarrow\infty$ asymptotics     
for the case $Q_1^2\,\simeq \,Q_2^2 \gg \mu_0^2$. We take $s$ to be much     
larger than the $Q_i^2$, but still within the region $I^+$ (\ref{plusext}):    
\beq\label{plusregion}    
1 \ll s/\tilde Q^2 \ll \tilde Q^2 /\mu_0^2.    
\eeq    
In this region the asymptotics is obtained from the asymptotic behavior of     
the Bessel function $I_0$     
\beq\label{as+}    
T^+(s\rightarrow    
\infty)\,=\,\frac{4\;\tau_{TT}}{\omega_0^2\,\,\sqrt{2\,\pi\,\omega_0\,\ln    
(s/\tilde Q^2)}}\,\left(\frac{s}{\tilde Q^2}\right )^{\omega_0}\,,    
\eeq    
and the result is entirely perturbative. When $s$ increases and     
eventually reaches the boarder line between $I^+$ and $I^-$:    
\beq    
s/\tilde Q^2 = \tilde Q^2/ \mu_0^2    
\eeq    
we have to switch to $T^-$. With a further increase of $s$, initially, the     
second term in (\ref{BS-12}) is not large and we can use the expansion     
of the Bessel function $I_2$ for small arguments. In the asymptotic region    
\beq    
\tilde Q^2/\mu_0^2 \ll s/\tilde Q^2       
\eeq    
the arguments of both Bessel functions are large. Expanding in both     
$\frac{1}{\ln  s/\tilde Q^2}$ and in the ratio     
$\frac{\ln \tilde Q^2/\mu_0^2}{\ln s/\tilde Q^2}$, we find that in    
(\ref{as-}) the leading terms of both Bessel function terms cancel against    
each other, and have to take into account first corrections. We obtain:    
\begin{eqnarray}\label{as-}     
T^-(s\rightarrow    
\infty)&=&\frac{8\;\tau_{TT}}{\omega_0^2\,\sqrt{2\,\pi}\,(\omega_0\,\ln    
(s/\tilde Q^2))^{3/2}}\,\left(\frac{s}{\tilde Q^2}\right    
)^{\omega_0}\,\times \nonumber \\ & &    
\\ & &\left[1\,+\,\omega_0\,\ln\frac{\tilde    
Q^2}{\mu_0^2}\,+\,\frac{\omega_0^2}{4}\,\ln^2\frac{\tilde    
Q^2}{\mu_0^2}\,+\,    
O\left(     
\frac{\omega_0^4\,\ln^4 \frac{\tilde Q^2}{\mu_0^2} }    
            {\ln \frac{s}{\tilde Q^2}}    
\right)     
\right]\,. \nonumber    
\end{eqnarray}    
    
It is interesting to compare (\ref{as+}) and (\ref{as-}): the power behavior     
in $s$ is the same in both regions. The difference lies in the     
preexponential factors: in the second region, we have a slightly stronger     
logarithmic suppression, and there is a logarithmic dependence upon the     
infrared scale $\mu_0^2$.

\subsection{DIS at low $x$}    
Another case of interest is deep inelastic scattering on an almost real photon    
at very small $x$. This corresponds to the limit     
\beq    
\mu_0^2 \approx Q_2^2 \ll Q_1^2 \ll s,    
\eeq    
and only the region $I^-$ applies ($Q\equiv Q_1$):    
\begin{eqnarray}\label{dis}   
T_{DIS}(Q^2,\,s)\,=\,\left [    
I_0\left(\omega_0\,\sqrt{\ln\frac{s}{    
Q^2}\,\ln\frac{s}{\mu_0^2}}\,\right)\,-\,\frac{\ln\frac{s}{Q^2}}    
{\ln\frac{s}{\mu_0^2}}\,    
I_2\left(\omega_0\,\sqrt{\ln\frac{s}{Q^2}\ln    
\frac{s}{\mu_0^2}}\right)\,-\,1\right]\,\frac{4\;\tau_{TT}}{\omega_0^2}\,.   
\nonumber \\    
\end{eqnarray}    
The Bjorken  $x$ is defined in a standard way: $x\,\equiv \,Q^2/s$. The    
flavor nonsinglet    
photon structure function is related to $T_{DIS}$ via    
\beq\label{fns}    
F_{NS}^{\gamma}(x,Q^2)\,    
=\,\frac{Q^2}{4\,\pi^2\,\alpha_{em}}\,\sigma^{\gamma^*\,\gamma}_{tot}\,    
=\,\frac{Q^2}{4\,\pi^2\,\alpha_{em}\,s}\,Im \,[T_{DIS}(Q^2,\,s)]\,   
\simeq\,\frac{x}{4\,\pi\,\alpha_{em}}\,    
\frac{\partial \,   
T_{DIS}(Q^2,\,s)}{\partial \ln\,s}\,.    
\eeq    
Being aware of the fact that in the DIS limit ($Q^2_2 \approx \mu_0^2$) the   
low $x$ saturation effects may be important \cite{IKMT} we will not   
discuss them here at all.   
   
We can consider two different asymptotic limits. The first one is    
$\ln\,1/x\,\gg\,\ln\,Q^2/\mu_0^2\,\gg\,1$.   
 In this limit the structure function    
becomes:    
\beq \label{fns1}    
F_{NS}^{\gamma}(x,Q^2)\,\simeq\,\left( \frac{1}{x} \right)^    
{-1\;+\; \omega_0}\,\,    
\frac{2\;\tau_{TT}\,(\,1\,+\,\omega_0\,\ln (Q^2/\mu_0^2)/4\,)^2}    
{\alpha_{em}\,\omega_0^2\,\sqrt{2\,\pi\,\omega_0}\,    
\ln^{3/2}(1/x) } \,\left(\frac{Q^2}{\mu^2_0}\right )^{\omega_0/2} \,.   
\eeq    
Eq. (\ref{fns1}) gives the Regge limit of the flavor nonsinglet structure    
function.    
Up to the preexponential factor    
this result agrees with the behavior of the flavor nonsinglet    
proton structure function found in ~\cite{Ry}.    
    
Another asymptotic limit to be considered is    
$1\,\ll\,\ln\,(1/x)\,\ll\,\ln\,(Q^2/\mu_0^2)$ leading to    
\beq \label{fns2}    
F_{NS}^{\gamma}(x,Q^2)\,\simeq\,x\,\,    
\frac{\tau_{TT}}    
{\alpha_{em}\,\pi\,\omega_0^2\,    
\sqrt{2\,\ln (1/x)\,\ln (Q^2/\mu_0^2)}} \,    
e^{\,\omega_0\,\sqrt{\ln (1/x)\,\ln (Q^2/\mu_0^2)}} \,.   
\eeq    
 Eq. (\ref{fns2}) comes from the asymptotic expansion of the first term   
in (\ref{dis}). The second term is subleading in this limit.     
Eq. (\ref{fns2}) corresponds to    
the double logarithmic limit of the DGLAP equation.    
    
\subsection{Scales}    
    
It is instructive to compare our results for the high energy behavior     
of quark-antiquark exchange in $\gamma^*\, \gamma^*$ scattering      
with those for gluon exchange, i.e. the LO BFKL Pomeron.     
For the latter it is well-known that, for sufficiently large photon     
virtualities and not too high energies, the internal transverse momenta     
are of the order of the photon virtualities and hence justify the use of     
perturbation theory (Fig. \ref{scalefig}a).    
When energy grows, diffusion in $\ln k^2$     
broadens the relevant region of internal transverse momenta, which has the     
shape of a ``cigar''. Its mean size is of the order $\sqrt{\ln s}$ 
and eventually     
reaches the infrared cutoff $\mu_0^2$. From now on, the BFKL amplitude -     
although infrared finite - becomes sensitive to infrared physics, and some     
modification due to nonperturbative physics has to be included.        
\FIGURE{   
\begin{tabular}{c c}   
\epsfig{file=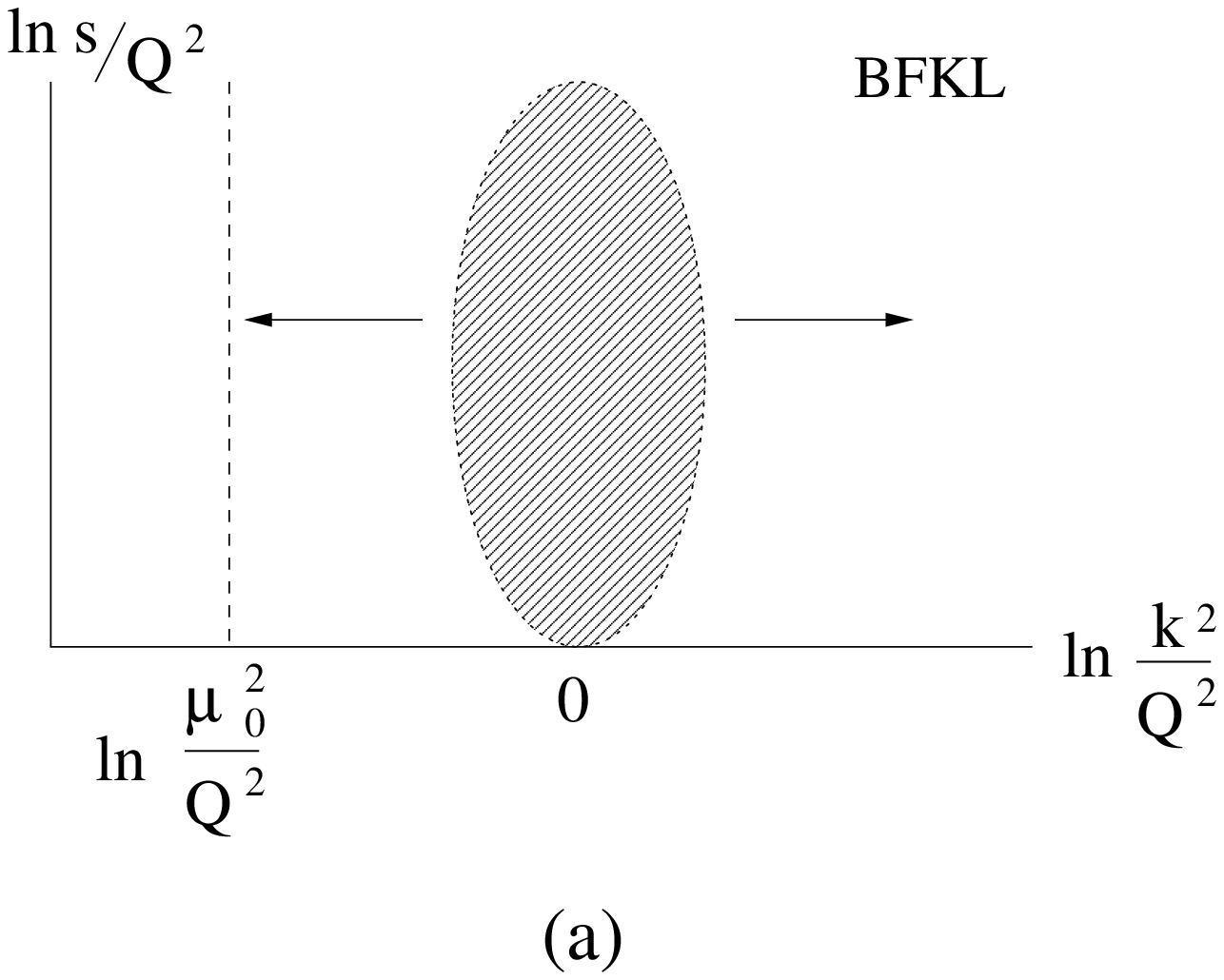,width=70mm}&   
\epsfig{file=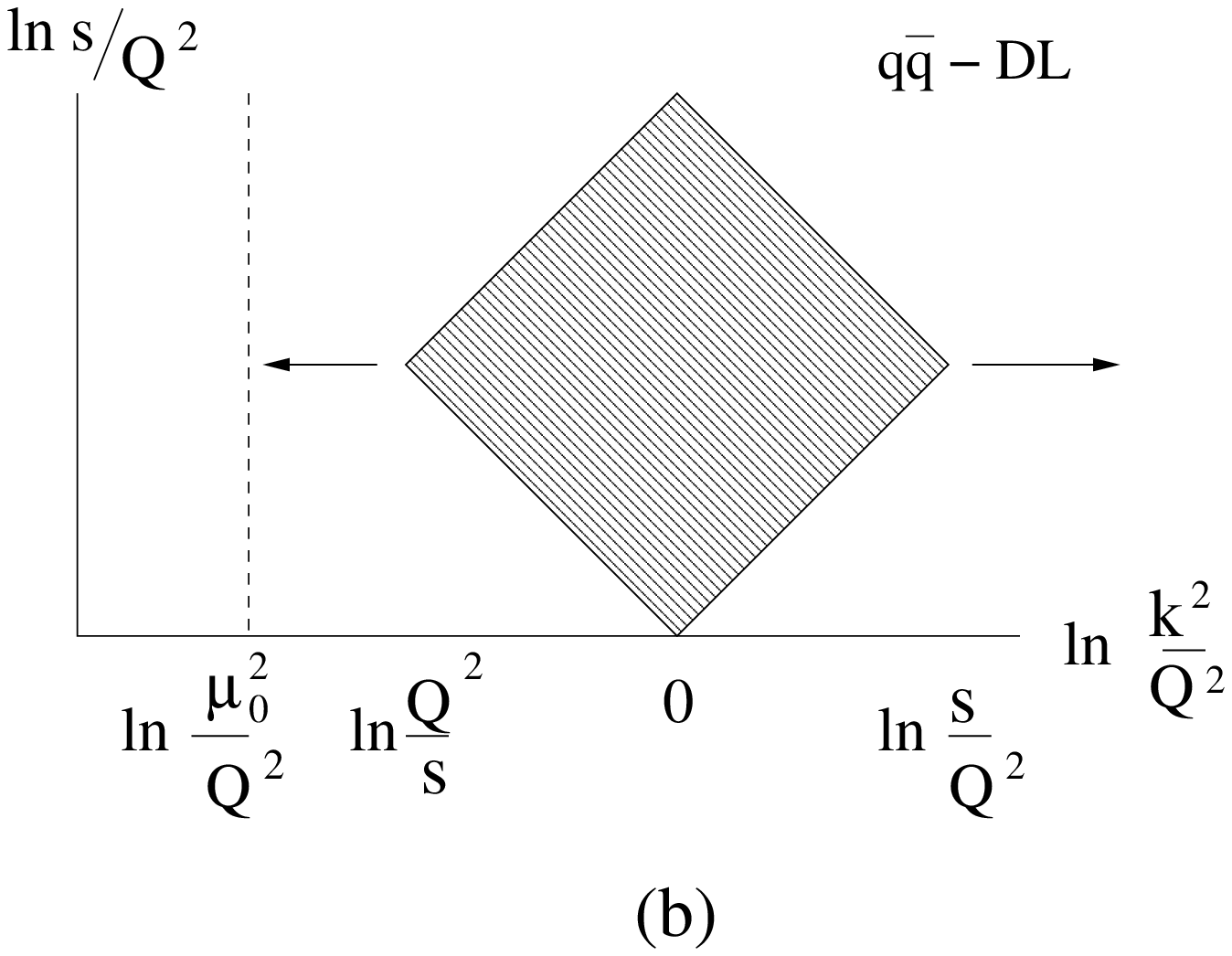,width=70mm}   
\end{tabular}   
  \caption{Energy dependence of the region   
of integration  in a) BFKL and b) quark ladder.}   
    \label{scalefig}}   
    
With the results of our analysis we now can make an analogous statement    
about quark-antiquark exchange. Since internal transverse momenta range  
between $max(Q^4/s,\,\mu_0^2) \,<\, k^2 \,< \,s$ or, equivalently   
\beq    
max(-\,\ln s/Q^2,\, \ln\,\mu_0^2/Q^2) \,\,<\,\, \ln\, k^2/Q^2\,\, <\,\, 
\ln \,s/Q^2\,,    
\eeq 
there is, for not too large energies, again, a limited region where 
transverse momenta stay above the infrared scale, and the use of    
perturbative QCD is justified. With increasing energy, the $k^2$-region  
expands and eventually hits the infrared cutoff $\mu_0^2$.  
From now on the high energy behavior starts to depend upon infrared physics  
and requires suitable modifications.  
 
In order to understand, in the fermion case, the region of internal  
integration in more detail, let us first return to our amplitude,  
$A^+$ (illustrated in Fig.\ref{BSfig}). 
It satisfies the two dimensional wave equation  
(\ref{BS+2}). It is instructive to introduce the new variables  
\begin{eqnarray} 
\label{time-space} 
t\,=\,\frac{1}{2}\,(\xi \,+\, \eta)\,=\,\ln \frac{\beta s}{\sqrt{k^2 Q^2}} 
\,;\,\,\,\,\,\,\,\,\,\,\,\,\,\,\,\,\,\,\,\,\,\,\,\,\,\,\,\,
z\,=\,\frac{1}{2}\,(\eta \,-\, \xi)\,=\,\frac{1}{2}\, \ln \frac{k^2}{Q^2},     
\end{eqnarray} 
which leads to the two-dimensional Klein-Gordon equation: 
\beq  
\label{KG} 
\left( \frac{\partial^2}{\partial t^2} \,-\, 
\frac{\partial^2}{\partial z^2} \,-\,  
4 \,\lambda \right)\,\, A^+\; =\; 0  
\eeq 
(here $i \sqrt{4 \,\lambda}$ plays the role of the mass). 
  \FIGURE{    
\epsfig{file=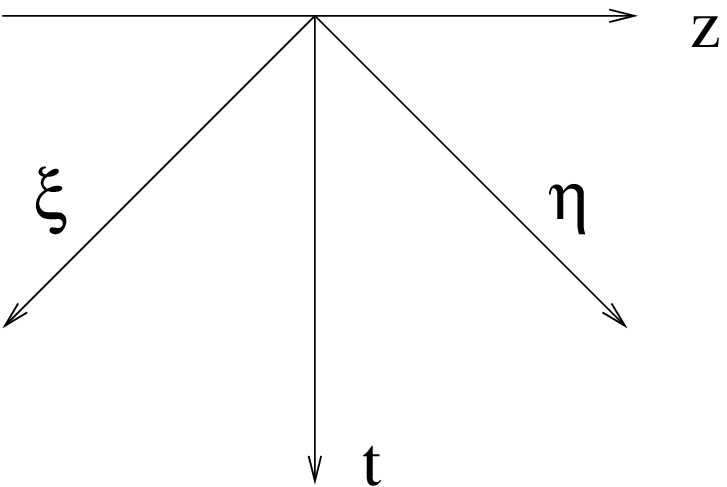,width=50mm}
  \caption{Light cone variables for the wave equation.}   
    \label{varfig}}  
As seen from (\ref{BS+1}), our amplitude  
satisfies the boundary condition $A^+(0,\,0)\,= \,1$, 
and its normal derivatives  
vanish on the `light cone' $\xi\,=\,0$ or $\eta\,=\,0$. 
Let us illustrate this  
in Fig. \ref{BSfig}: the upper end of the ladder we associate with  
the values $\beta \,\approx\, Q^2/s$ and $k^2\, \approx \,Q^2$, 
which translates  
into $\xi\,=\,\eta\,=\,0$ or $t\,=\,z\,=\,0$. 
At the lower end, when coupling to the external  
photon, we put $\beta\,=\,1$, $k^2\,=\,Q^2$, which is equivalent to  
$\xi\, =\, \eta \,=\, \ln \,s/Q^2$ or $t\,=\,\ln \,s/Q^2$, ~
$z\,=\,0$.  
Correspondingly, in Fig.\ref{varfig} the $t$-axis points downwards: 
our ladder  
diagrams for $\gamma^*\, \gamma^*$ describe the evolution (in $t$) from  
the initial point $(t_i,\,z_i)\,=\,(0,\,0)$ (upper photon) to  
the final point $(t_f,\,z_f)\,=\,(\ln \,s/Q^2,\,0)$ (lower photon).  
As seen from (\ref{BS+1}), all paths lie in the region of positive  
$\xi$ and $\eta$, i.e. they stay inside the square described by  
$0 \,<\, \xi\, <\, \xi_f\,=\, \ln \,s/Q^2$, ~
$0 \,< \,\eta\, < \,\eta_f\,=\, \ln \,s/Q^2$. 
Note the width in $z$: it starts from  
zero, increases up to its maximal value $\ln \,s/Q^2$, and finally  
it shrinks down to zero again.   
 
Let us confront this with the BFKL Pomeron: in Fig.\ref{scalefig}b  
we have, once more, drawn the square which illustrates the internal region  
of integration. Apart from the difference in shape (``diamond'' versus 
``cigar''), 
the most notable difference is the width in $\ln k^2$. In the fermion case  
it grows proportional to $\ln \,s/Q^2 $, i.e stronger than in the  
BFKL case where $\ln\, k^2$ grows as $\sqrt{\ln \,s/Q^2}$:  
the BFKL diffusion is replaced by a linear growth in the $z$-direction.   
Also the definition of `internal rapidity' is different: in the BFKL case  
the vertical axis can be labeled simply by $\ln \,1/\beta$, whereas in the  
fermion case our variable is 
$t\,=\,\ln \,(\beta \,s / \sqrt{k^2\, Q^2})$ (in both  
cases, the total length grows proportional to $\ln s$).                    
    
\section{InfraRed Evolution Equation}

In this section we rederive our previous result from the IREE.      
We define the partial wave through the ansatz::    
\beq\label{pw}   
T(Q_1^2,\,Q_2^2,\,s)\,=\,\int    
\frac{d\omega}{2\,\pi\,i}\,\left(\frac{s}{\tilde\mu^2}\right)^\omega\,
\tilde    
T(Q_1^2/\tilde\mu^2,\,Q_2^2/\tilde\mu^2,\,\omega)\,.  
\eeq    
    
For a more detailed description of the    
method we refer the reader to the original work \cite{KiLi} as well    
as to some applications in Ref. \cite{BER2,Ry,BER1,KMSS}.     
The scale $\tilde \mu$    
is introduced as an auxiliary infrared cutoff parameter, which, a priori, 
does    
not have to coincide with the infrared cutoff scale of    
nonperturbative physics, $\mu_0$. Before applying the IREE formalism to    
$\gamma^* \,\gamma^*$ scattering, we have to discuss its applicability.    
Based upon our previous analysis, we    
have to distinguish between two cases. For any t-channel quark line    
we have the condition that $\mu_0^2 \,<\, s\, \alpha\, \beta \,= \,k^2$    
which, in the    
derivation of the IREE, will be replaced by $\tilde \mu^2 \,<\, s\,   
 \alpha\, \beta\, =\,k^2$.    
On the other hand, we have the constraints that $Q_1^2/s < \beta$, and    
$Q_2^2/s < \alpha$. Combining these conditions we have    
$max(\tilde \mu^2, \,Q_1^2 \,Q_2^2/s) \,<\, s\,\alpha \,\beta\, =\,k^2$,    
and this distinction (after identifying    
the auxiliary parameter $\tilde \mu$ with the physical infrared cutoff    
$\mu_0$) leads to the    
the two regions, $I^+$ and $I^-$. In the latter case, all internal    
transverse momenta are cutoff by $\tilde \mu$, and this is the case where    
the IREE directly applies. In the former case, the internal momenta    
do not reach the value $\tilde \mu$, i.e. the amplitude is independent    
of $\tilde \mu$. This case, therefore, has to be studied separately.    
    
Let us begin with the genuine IREE region, $Q_1^2 \,Q_2^2/s \,<\,    
\tilde \mu^2$.    
The IREE for the    
amplitude $T$ is obtained by differentiating (\ref{pw}) with respect to 
$\ln  \tilde\mu^2$.     
In $\omega$ space this essentially corresponds to cutting    
 of the diagram on two parts (see Fig \ref{IREEfig}).    
\FIGURE{   
\epsfig{file=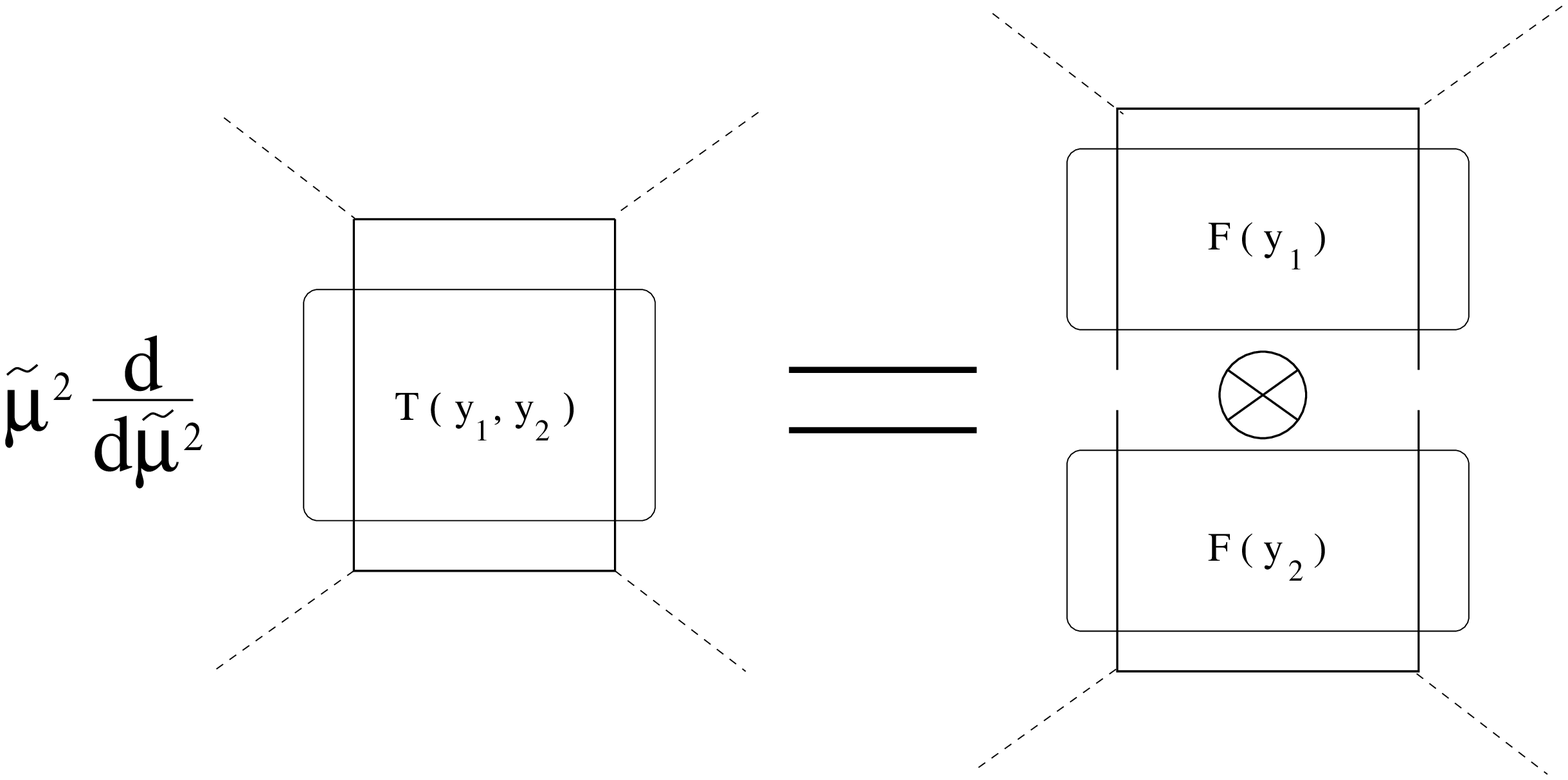,width=142mm}   
 \caption{The IRRE for $\gamma^*\,\gamma^*$.}   
    \label{IREEfig}}   
   
The equation reads    
\beq\label{IREE}   
\omega\,\tilde    
T^-\,\,+\,\,\frac{\partial \tilde T^-}{\partial y_1}\,\, +\,\,\frac{\partial    
\tilde T^-}{\partial y_2}\,\,=\,\,2\,N_c\,F_{nc}   
\,\,\frac{F(\omega, y_1)\,\,F(\omega,    
y_2)}{8\,\pi^2}   
\eeq    
with $y_{1,2}\,\equiv\,\ln (Q_{1,2}^2/\tilde\mu^2)$,      
\beq\label{F}    
F(\omega,\,y)\,=\,\frac{C_0}{g^2\,C_F}\,f_0(\omega)\,e^{\,-(\omega-f_0/8\pi^2)\,y}    
\,\,\,\,\,\,\,\,\,\,\,\,\,\,\,\,for\,\,\,\,\,\,\,\,y \ge 0\,,  
\eeq    
and $C_0\,=\,4\,\pi\,\alpha_{em}\,\epsilon (A) \,\cdot\,\epsilon (A^\prime)$.    
    
The function $f_0$ was introduced in Ref. \cite{KiLi}     
\beq\label{f0}    
f_0(\omega)\,=\,4\,\pi^2\,(\omega\,-\,\sqrt{\omega^2-\omega_0^2})    
\eeq     
with the property:    
\beq \label{prop}    
f_0 \,(\omega\, -\, \frac{f_0}{8\,\pi^2})\, =\, 2\,\pi^2 \,\omega_0^2\,.    
\eeq    
The function $F(\omega,\, y)$ is a solution of the IREE for a    
quark-photon amplitude \cite{KiLi, Ry}. The coefficient $C_0$    
contains information about coupling of the photon to the ladder.   
Note the appearance in (\ref{IREE})    
of the factor $ 2\,N_c\,F_{nc}$ which incorporates    
a proper normalization   
for the case of $\gamma^*\,\gamma^*$.   
    
We should impose  boundary conditions for the case    
when only one photon is (nearly) on shell.  For the boundary    
$y_1\,\ge\,y_2\,=\,0$ it reads:    
\beq\label{bc1IREE}    
\omega\,\tilde T^-(y_1,\,y_2\,=\,0)\,\,+\,\,    
\frac{\partial \tilde T^-(y_1,\,y_2\,=\,0)}{\partial y_1}    
\,\,=   
\,\,2\,N_c\,F_{nc}\,\frac{F(\omega, y_1)\,F(\omega, y_2\,=\,0)}{8\,\pi^2}\,.    
\eeq    
A similar condition is imposed at the second boundary $y_2\,\ge\,y_1\,=\,0$.    
When both photons are (nearly) on shell we have an additional equation    
\beq\label{bcIREE}    
\omega\,\tilde T^-(y_1\,=\,y_2\,=\,0)\,\,=\,\, 2\,N_c\,F_{nc}\,   
\frac{F^2(\omega, y\,=\,0)}{8\,\pi^2}. \eeq    
    
We will find now a solution of    
Eq. (\ref{IREE}) with the boundary conditions    
(\ref{bc1IREE}) and (\ref{bcIREE}). Note that the solution to be found is    
unique.    
    
A general solution of  the inhomogeneous equation    
(\ref{IREE}) can be found as a sum of a particular solution ($\tilde T_{ps}$)    
and a solution of the homogeneous equation ($\tilde T_{he}$), such that    
$\tilde T^-\,=\,\tilde T_{ps}\,+\,\tilde T_{he}$.    
We will search for $\tilde T_{ps}$ using the following ansatz,    
which is motivated    
by the  right hand side of (\ref{IREE}):    
\beq\label{anTps}    
\tilde T_{ps}(\omega)\,=\,    
\tilde T_{ps}^0(\omega)\,e^{\,-(\omega\,-\,f_0/8\pi^2)\,(y_1\,+\,y_2)}\,.    
\eeq    
Substituting it to (\ref{IREE}) we find $\tilde T_{ps}^0(\omega)$:    
\beq\label{Tps}    
\tilde T_{ps}(\omega)\,=\,-\,\kappa    
\frac{f_0^2}{\omega\,-\,f_0/4\,\pi^2}\,    
e^{\,-(\,\omega\,-\,f_0/8\pi^2)\,(y_1\,+\,y_2)}    
\eeq    
with    
\beq\label{kappa}    
\kappa\,=\,\frac{C_0(A)\,C_0(B)\,2\,N_c\,F_{nc}}{8\,\pi^2\,g^4\,C_F^2}\,   
=\,\frac{\tau_{TT}}{g^4\,C_F^2}    
\,=\,\frac{\tau_{TT}}{4\,\pi^4\,\omega_0^4}\,.   
\eeq

Our goal now is to find $\tilde T_{he}$ which satisfies the homogeneous    
version of Eq.(\ref{IREE}). Define two new variables:    
$w\,=\,y_1\,+\,y_2$ and $u\,=\,y_1\,-\,y_2$. Then    
\beq\label{The}    
\omega\,\tilde T_{he}(\omega)\,+\,2\,\frac{\partial \tilde T_{he}}    
{\partial w}\,=\,0    
\eeq    
with the solution    
\beq\label{Thes}    
\tilde T_{he}(\omega,u,w)\,=\,    
\tilde T^0_{he}(\omega,\,u)\,e^{-\omega\,w/2}\,=\,    
\tilde T^0_{he}(\omega,\,y_1\,-\,y_2)\,e^{-\omega\,(y_1\,+\,y_2)/2}.    
\eeq    
The boundary conditions (\ref{bc1IREE}) and (\ref{bcIREE}) will be used    
in order to find $ \tilde T^0_{he}$.  When    
$\tilde T^-\,= \,\tilde T_{ps}\,+\,\tilde T_{he}$ is    
substituted in Eq. (\ref{bc1IREE}) we get ($y_2=0$)    
\beq\label{The0}    
\frac{\omega}{2}\,\tilde T^0_{he}\,+\,\frac{\partial \tilde T^0_{he}}    
{\partial y_1}\,=\,\kappa\,\frac{f_0^2\,(\omega\,-\,f_0/8\,\pi^2)}    
{\omega\,-\,f_0/4\,\pi^2}\,\,e^{\,-\,(\omega\,-\,f_0/4\,\pi^2)\,y_1/2}\,.    
\eeq    
With the help of (\ref{prop}) the rhs can be simplified:    
\beq    
\frac{\omega}{2}\,\tilde T^0_{he}\,+\,\frac{\partial \tilde T^0_{he}}    
{\partial y_1}\,=    
\kappa\, \frac{2 \,\pi^2\, \omega_0^2\, f_0}{\omega\,-\,f_0/4\,\pi^2}    
\,\,e^{\,-\,(\omega\,-\,f_0/4\,\pi^2)\,y_1/2}\,.    
\eeq    
Eq. (\ref{The0}) has a solution which, again, is given by a sum of a    
particular solution ($R_{ps}$)  and a solution of the homogeneous equation    
($R_{he}$), such that $\tilde T^0_{he}\,=\,R_{ps}\,+\,R_{he}$. The particular    
solution is    
\beq\label{Rps}    
R_{ps}(\omega,\,y_1)\,=\,    
\kappa\,\frac{16\, \pi^4 \,\omega_0^2}    
{\omega\,-\,f_0/4\,\pi^2}\,\,e^{\,-\,(\omega\,-\,f_0/4\,\pi^2)\,y_1/2}\,,    
\eeq    
while the solution of the homogeneous equation is    
\beq\label{Rhe}    
R_{he}(\omega,\,y_1)\,=\,R_{he}^0\,e^{\,-\omega\,y_1/2}\,.    
\eeq    
Finally we use the boundary condition (\ref{bcIREE}) in order to determine    
$R_{he}^0$:    
\beq\label{Rhe0}    
R_{he}^0(\omega)\,=\,-\kappa\,\frac{16\, \pi^4\, \omega_0^2}    
{\omega}\,.    
\eeq

Combining all the results together we find the desired solution:    
    
\begin{eqnarray}\label{solIREE} \tilde    
T^-(\omega)\,&=&\,-\,\kappa \,   
\frac{f_0^2}{\omega\,-\,f_0/4\,\pi^2}\,    
e^{\,-(\omega\,-\,f_0/8\pi^2)\,(y_1\,+\,y_2)}\,\,\,+    
\nonumber \\ & &\\    
& &16\,\pi^4\, \kappa\, \omega_0^2 \,   
e^{\,-\,y_1\,\omega}\,\left[\,    
\frac{1}    
{(\omega\,-\,f_0/4\,\pi^2)}\,e^{\,-(y_2\,-\,y_1)\,f_0/8\,\pi^2}\,-\,    
\frac{1}{\omega}\,    
\right]\,,\nonumber    
\end{eqnarray}    
which we can also write as    
\begin{eqnarray}\label{solIREE2}    
\tilde T^-(\omega) &=& - \,16\, \pi^4\, \kappa \,   
\frac{(\omega \,-\, \sqrt{\omega^2\, -\,    
\omega_0^2})^2}{\sqrt{\omega^2 \,-\, \omega_0^2}}\,    
e^{\,-(\omega\,-\,f_0/8\,\pi^2)\,(y_1\,+\,y_2)}    
\nonumber\\    
&+& 16\, \pi^4 \,\kappa\, \omega_0^2 \,   
\left[ \frac{e^{\,-\,\omega (y_1\,+\,y_2)/2}    
\,\,e^{\,(y_2\,-\,y_1) \sqrt{\omega^2 \,-\, \omega_0^2}/2}}    
{\sqrt{\omega^2 \,-\, \omega_0^2}} \,-\,    
\frac{e^{\,-\,y_1\,\omega}}{\omega} \right].    
\end{eqnarray}    
In (\ref{solIREE}) and (\ref{solIREE2}), the first term on the rhs    
is the particular solution of Eq. (\ref{IREE}), whereas the second term    
is a solution of the homogeneous equation.    
Since we used the boundary condition (\ref{bc1IREE}),    
the solution (\ref{solIREE}) (or (\ref{solIREE2})) is valid for $y_1\,>\,y_2$.    
Would we have started with the boundary conditions at $y_1=0$,    
we would have ended up with (\ref{solIREE}) or (\ref{solIREE2}),    
but with $y_1$ and $y_2$ being interchanged.    
    
At first sight, the distinction between the two cases    
$y_1 \,<\, y_2$ and $y_2\, <\, y_1$ might look somewhat strange.        
Returning with our result into (\ref{pw}), putting    
$\tilde \mu^2 \,=\, \mu_0^2$, we reproduce exactly the expression (\ref{Tamp-})   
for the amplitude $T^-$. It is not difficult to    
identify (after integration over $\omega$) the first term on the rhs of    
(\ref{solIREE2}) with the second term on the rhs of (\ref{Tamp-}) (the Bessel    
function $I_2$), whereas the ``nonsymmetric''    
second term equals $I_0 \,-\, 1$.     
The final result is fully symmetric with respect   
to $y_1\,\leftrightarrow\,y_2$ interchange.   
The apparent puzzle is resolved because   
of the two-sheeted structure of the $\omega$ plane for    
 the partial wave representation of the   
Bessel function $I_0$ (see Appendix B).

Finally, we return to the other region, $\tilde \mu^2 \,<\, Q_1^2\, Q_2^2/s$.    
As we have argued above, in this region the amplitude does not depend upon    
$\tilde \mu^2$. Starting from the ansatz (\ref{pw}), we have to require    
that the total derivative with respect to $\tilde \mu^2$ vanishes, i.e.    
\beq    
\omega\,\tilde    
T^+\,+\,\frac{\partial \tilde T^+}{\partial y_1}\,\, +\,\,\frac{\partial    
\tilde T^+}{\partial y_2}\,\,=\,\,0\,.      
\eeq     
A solution to this equation has to be of the form (\ref{Thes}), where the    
dependence upon the difference $y_1\,-\,y_2$ has to be fixed by boundary    
conditions. We require that, for $\tilde \mu^2\, =\, Q_1^2\, Q_2^2/s$    
the solution matches the solution (\ref{solIREE2}), found for the other    
region, $I^-$. Since, on the rhs of (\ref{solIREE2}), the first term     
after integration over $\omega$ is proportional to the Bessel    
function $I_2(\omega_0\, \sqrt{\ln (s \tilde\mu^2/\tilde Q^2) \,   
\ln (s/\tilde \mu^2)})$ and vanishes for $s \tilde\mu^2/\tilde Q^2 \,=\,1$,    
we are left with the second term only and our result is          
\beq    
\tilde T^+(\omega) \,=\,    
16 \,\pi^4 \,\kappa \,\omega_0^2 \,   
\left[ \frac{e^{\,-\,\omega \,(y_1\,+\,y_2)/2} \,\,   
e^{\,(y_2\,-\,y_1)\, \sqrt{\omega^2 \,-\, \omega_0^2}/2}}    
{\sqrt{\omega^2 \,-\, \omega_0^2}} \,-\,    
\frac{e^{\,-\,y_1\,\omega}}{\omega} \right].    
\eeq       
After integration over $\omega$, this becomes $I_0 \,-\,1$,    
in agreement with        
(\ref{Tamp+}).    
    
In summary, we have been able to derive, within the IREE approach, both our    
solutions $T^-$ and $T^+$. This result allows for the possibility of 
studying    
also non-ladder diagrams which appear in odd signature exchange 
amplitudes.       
    
 \section{Reggeon Green's function}   
   
A third method of describing the double logarithmic high energy behavior    
of quark-antiquark exchange has been developed in Ref. \cite{Kirschner}.   
Kirschner has shown that the double logarithmic situation can be    
described in  the same way as the BFKL Pomeron, namely by a two-reggeon    
Green's function with a kernel which is conformal invariant and holomorphic    
separable. In particular, in the integral equation of the reggeon Green's    
function the transverse momentum integration has no explicit infrared or    
ultraviolet cutoff; therefore it has to be compared with our result for    
$\mu_0=0$. In the following we will show that, starting from Kirschner's    
Green's function, we arrive at the same result as in section 4.           

The reggeon Green function in Ref. \cite{Kirschner}    
has been obtained by deriving and solving a linear BFKL-type equation    
for the quark  exchange amplitude \cite{KK}. The Green function has the form    
\beq    
G(k_\perp\,,\,\bar k_\perp\,,\,\omega)\,=\,\frac{1}{2\,\pi^2}    
\int_{-\infty}^\infty d\,\nu\,\sum_{n=-\infty}^\infty    
\frac{|k_\perp^2|^{-\,1\,+\,i\,\nu}\left(\frac{k_\perp}{|k_\perp|}\right)^n\,    
|\bar k_\perp^2|^{-\,1\,-\,i\,\nu}   
\left(\frac{\bar k_\perp}{|\bar k_\perp|}\right)^{-n}}    
{\omega\,-\,\omega_0^2\,\Omega(\omega,\,\nu,\,n)/4}    
\label{green}    
\eeq    
with    
\begin{eqnarray}    
\Omega(\omega,\,\nu,\,n)\,=\,4\,\psi(1)\,-\,\psi(-\,i\,\nu\,+\,   
\frac{|n|}{2}\,+\,\frac{\omega}{2})\,-\,    
\psi(i\,\nu\,+\,\frac{|n|}{2}\,+\,\frac{\omega}{2})\,-\, \nonumber \\    
\psi(1\,-\,i\,\nu\,-\,\frac{|n|}{2}\,+\,\frac{\omega}{2})\,-\,    
\psi(1\,+\,i\,\nu\,-\,\frac{|n|}{2}\,+\,\frac{\omega}{2})    
\label{Om}    
\end{eqnarray}    
and $\psi(z)\,=\,\frac{d}{d\,z}\,\ln \Gamma(z)$.    
The function $G$ is normalized such that at the Born level it equals    
$G_0=\frac{1}{\omega\,k_\perp^2}\,\delta^2(k_\perp\,-\,\bar k_\perp)$,    
which is  a propagator of two quark lines.

At high energies the leading contribution comes from $n\,=\,0$.    
 After expansion of $\Omega$ for small    
values of $\omega$ and $\nu$ the Green functions reduces to    
    
\beq    
G(k_\perp\,,\,\bar k_\perp\,,\,\omega)\,=\,\frac{1}{2\,\pi^2}    
\frac{1}{k_\perp^2\,\bar k_\perp^2}\int_{-\infty}^\infty d\,\nu\,    
\frac{\omega^2\,+\,4\,\nu^2}
{\omega\,(\omega^2\,-\,\omega_0^2\,+\,4\,\nu^2)}\,    
\left(\frac{k^2_\perp}{\bar k_\perp^2}\right)^{i\,\nu}    
\label{green1}    
\eeq    
    
The $\gamma^*\,\gamma^*$ scattering amplitude can be obtained from   
$G$  by integrating over momenta $k$    
in the low and upper blocks in Fig. \ref{Grfig}.      
Note that the propagators of the t-channel quarks ($1/k^2$)    
are already included  in $G$. In Ref. \cite{Kirschner} they were assumed    
to be purely transverse implying that   
\beq\label{ab}   
\alpha\,\beta\,s\,\ll\,k_\perp^2    
\,\,\,\,\,\,\,\,\,\,\,\,\,\,\,\,\,\,\,\,\,   
and \,\,\,\,\,\,\,\,\,\,\,\,\,\,\,\,\,\,\,\,    
\bar\alpha\,\bar\beta\,s\,\ll\,\bar k_\perp^2\,.   
\eeq    
The $\bar \alpha$ and $\beta$ integrations pick up the  
poles of the $s$-channel  
quark propagators.  
 The only trace of these integrations is in the ordering (\ref{ord})   
in $\bar k^2/\bar\beta$ (and in $k^2/\alpha$).    
This ordering leads to the following limits for $\beta$ and $\bar\alpha$:    
$$\,\alpha_{max}\,=\,min\{1\,,k_\perp^2/Q_2^2\} \,    
\,\,\,\,\,\,\,\,\,\,\,\,\,\,\,\,\,\,\,\,\,\,\,\,\,    
\bar \beta_{max}\,=\,min\{1\,,\bar k_\perp^2/Q_1^2\}.$$    
The scattering amplitude has the form   
\begin{eqnarray}    
T_G^{\gamma^*\,\gamma^*}(Q_1^2,\,Q^2_2,\,s)\,=\,\tau_{TT}\,   
\int d^2\,k_\perp    
\int_0^{\alpha_{max}}\frac{d\alpha}{\alpha}   
\int d^2\,\bar k_\perp    
\int_0^{\bar \beta_{max}}\frac{d\bar\beta}{\bar\beta}     
\nonumber \\ \times\,   
\int \frac{d\,\omega}{2\,\pi\,i}\,    
\left (\frac{\alpha\,\bar\beta\,s}    
{|k_\perp|\,|\bar k_\perp|}\right)^\omega\,\omega\,    
G(k_\perp,\,\bar k_\perp,\,\omega) \,.   
\label{sr}    
\end{eqnarray}   
The overall coefficient $\tau_{TT}$ as well as the    
factor $\omega$ in front of $G$ account for the proper    
normalization  and can be deduced using the Born approximation.   
Note that, following the derivation of $G$ in \cite{Kirschner}, we   
assume that any potential infrared and ultraviolet divergences stemming   
from the $k_\perp$ integrations are regularized by the $\omega$   
dependent factor. This is why the $\omega$ integration should be done   
only after the $k_\perp$ integrations are performed.   
   
After the $\alpha$ and $\bar\beta$ integrations we obtain:     
\beq    
T_G^{\gamma^*\,\gamma^*}\,=\,\tau_{TT}\,   
\int d^2\,k_\perp    
\int d^2\,\bar k_\perp    
\int \frac{d\,\omega}{2\,\pi\,i\,\omega}\,    
\left (\frac{\alpha_{max}\,\bar\beta_{max}\,s}    
{|k_\perp|\,|\bar k_\perp|}\right)^\omega\,    
G(k_\perp,\,\bar k_\perp,\,\omega)\,.    
\label{regge}    
\eeq    
It is worth to notice that   
 the $\gamma^*\,\gamma^*$ scattering amplitude can be brought to the form   
in which $G$ is sandwiched between the photon impact factors:     
\begin{eqnarray}\label{TG}   
T_G^{\gamma^*\,\gamma^*}\,=\,N_c\,F_{nc}\,\int    
\frac{d^2\,k_\perp}{2\,\pi}\int   
\frac{d^2\,\bar k_\perp}{2\,\pi}    
\int \frac{d\,\omega}{2\,\pi\,i\,\omega}\, \left (\frac{s}    
{|k_\perp|\,|\bar k_\perp|}\right)^\omega\, \nonumber \\  \\ \times\,   
 \Phi(Q_1^2,\, k_\perp^2,\,\omega)\,\,   
G(k_\perp,\,\bar k_\perp,\,\omega) \,\,\Phi(Q_2^2,\, \bar k_\perp^2,\,\omega).   
\nonumber   
\end{eqnarray}   
The photon impact factor has the form   
\beq\label{if}   
\Phi(Q^2,\, k_\perp^2,\,\omega)\,=\,C_0\,   
\left[\,\left(\frac{k_\perp^2}{Q^2}\right)^\omega   
\,\,\Theta\left(1\,-\,\frac{k_\perp^2}{Q^2}\right)\,\,+\,\,   
\Theta\left(\frac{k_\perp^2}{Q^2}\,-\,1\right)\,\right]\,.   
\eeq   
with $C_0$ defined in the previous section.   
 Integrating in (\ref{regge})     
over $k_\perp\,,\bar k_\perp$ we get    
\begin{eqnarray}\label{kint}    
T_G^{\gamma^*\,\gamma^*}\,=\,    
\frac{8\,\tau_{TT}}{\pi}\,\int \frac{d\,\omega}{2\,\pi\,i}\,    
\int_{-\infty}^\infty d\,\nu\,    
\frac{1}{(\omega^2\,+\,4\,\nu^2)\,    
(\omega^2\,-\,\omega_0^2\,+\,4\,\nu^2)}\,\left(\frac{s}{\tilde Q^2}\right)    
^\omega \,\left(\frac{Q_1^2}{Q_2^2}\right)^{i\,\nu} \,.   
\end{eqnarray}    
The $\nu$ integration leads to    
\beq    
T_G^{\gamma^*\,\gamma^*}\,=\,    
\frac{4\,\tau_{TT}}{\omega_0^2}\,\int \frac{d\,\omega}{2\,\pi\,i}\,    
\left (\frac{s}{\tilde Q^2}\right)^\omega\, \left[\,   
\left(\frac{Q_1^2}{Q_2^2}\right)^{-\sqrt{\omega^2-\omega_0^2}/2}\,   
\frac{1}{\sqrt{\omega^2\,- \,\omega_0^2}} \,-\,\frac{1}{\omega}\,   
\left(\frac{Q_1^2}{Q_2^2}\right)^{-\omega/2}\,\right]\,.   
\label{regge1}    
\eeq    
Finally, after the $\omega$ integration we arrive at:   
\beq    
T_G^{\gamma^*\,\gamma^*}\,=\,\frac{4\,\tau_{TT}}{\omega_0^2}\,     
\left[\,I_0\,\left(\omega_0\,\sqrt{\ln \frac{s}{Q_1^2}\,\ln \frac{s}{Q_2^2}}    
\,\right)\,\,-\,\,1\,\right] \,.   
\label{sregge}    
\eeq    
  
The expression (\ref{sregge}) coincides exactly with the amplitude   
$T^+$ (\ref{Tamp+}), with the restriction $\mu_0^2 < Q^4/s$ being removed.      
Formally we can say that the Green's function approach should be    
compared to our previous treatment (in particular, the linear equation    
in section 4) with $\mu_0^2=0$ being put to zero. A striking difference    
between the two treatments lies in the integration over transverse momenta.   
From the discussion in section 4 it follows that all transverse momenta   
inside the ladder lie in the region $Q^4/s\,< \,k^2 \,<\,s$. 
In the integral equation    
for the reggeon Green's function \cite{Kirschner}, on the other hand, the      
Regge energy factor $\left(s/k_{\perp} k_{\perp}'\right)^{\omega}$    
serves as a regulator, and the transverse momentum integration can be     
extended down to zero and up to infinity.    
      \FIGURE{   
\epsfig{file=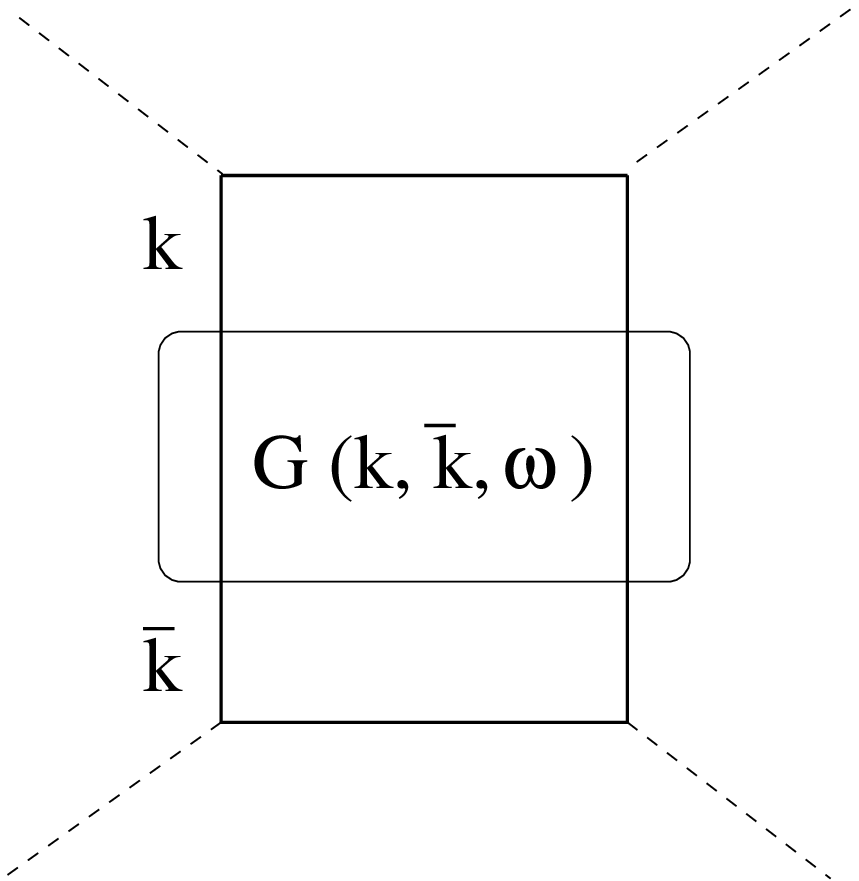,width=55mm}   
 \caption{ $\gamma^*\,\gamma^*$ via reggeon Green function.}   
 \label{Grfig}}
An easy way to see how this works is the following. 
Return to our two-loop integral in section 3 and compare the    
lower cell, $k_2$, with the $\bar k$ integration in (\ref{sr}). The    
integration region is illustrated in Fig. \ref{int2}a:   
interchanging the order    
of integration over $\beta_2$, $k_2$, we see that the $k_2$ integration    
goes from $Q^4/s$ to $s$ and splits into two pieces. 
For $Q^4/s \,<\, k_2^2\, <\,Q^2$, the $\beta_2$ integral   
extends from $Q^2/s$ to $k_2^2/Q^2$, and for $Q^2 \,<\, k_2^2\, <\, s$    
we have $k_2^2/s \,<\, \beta_2\, <\, 1$. 
If we now replace, in Fig. \ref{int2}a, the upper    
cell by the reggeon Green's function, including the energy factor    
$(\beta_2 s/\sqrt{Q^2 k_2^2})^{\omega}$, we see that, as long as    
Re $\omega \,>\,0$, we can extend the $\beta_2$ integral down to zero and    
enlarge the transverse momentum integral to the whole 
interval $[0,\infty]$.    
 
Nevertheless, from the discussion in section 3 we know that the double   
logarithmic high energy behavior comes from the restricted region       
$Q^4/s\,<\, k_2^2\, <\,s$ (more precisely, from the shaded region in  
Fig. \ref{scalefig}b). In the Green's function approach, this restriction  
of internal integration appears as soon as we replace the  
integrand of (\ref{green}) by the $n\,=\,0$, small-$\nu$ and
 $\omega$ approximation, 
(\ref{green1}). Before the making the approximation, there is no  
restriction on paths connecting, in Fig. \ref{scalefig}b, 
the starting point at the top  
with the end point at the bottom. In particular, there will be paths outside  
the marked square. Turning to the leading high energy behavior and doing the 
$n=0$, small-$\nu$ and $\omega$ approximations,      
it is easy to verify that (\ref{green1}), when  
coupled to the upper photon, is identical to our $A^+$ function\footnote{It is 
important to note that, in (\ref{green1}), the contours  
of integration of the $\nu$ and $\omega$ integrals, are not independent of 
each other. For example, one first does the $\nu$ integral (at fixed  
$Re\,\omega\,>\,\omega_0$), then closes the remaining $\omega$ contour around the  
cut at $-\omega_0\, <\, \omega \,<\, \omega_0$}: is satisfies  
the two-dimensional wave equation, and the discussion of section 4.5 applies. 
In this sense one might say that approximation which leads from   
(\ref{green}) to (\ref{green1}) plays the same role as, in the BFKL case,  
the saddle point approximation.

\section{Numerical estimates}    
    
The final goal of this project should be a confrontation    
of the obtained results  with the LEP data. We are not ready    
yet to produce such a comparison since we still miss    
a significant theoretical contribution arising from    
the flavor nonsinglet exchange. In this section we will present    
some first numerical estimates related to the resummation of the 
quark ladder.    
    
Using Eq. (\ref{sigma})    
the flavor nonsinglet contribution to $\sigma^{\gamma^*\,\gamma^*}_{tot}$    
can be computed from the elastic amplitude (\ref{Tamp+}) and (\ref{Tamp-}).    
In our numerical estimates we will drop the flavor factor $F_{ns}$:    
the missing flavor singlet piece will be estimated to have the same    
functional form as the nonsinglet piece, and (\ref{sigma}) - with the    
factor $F_{ns}$ being replaced by $\sum \,e_q^4$    
- is assumed to represent the sum of    
flavor singlet plus flavor nonsinglet.        
    
First let us demonstrate numerically the effect of the two kinematical    
regions appearing in  (\ref{Tamp+}) and (\ref{Tamp-}). 
Fig. \ref{sigfig} shows    
the $\gamma^*\,\gamma^*$ cross section as a function of rapidity    
$Y\,=\,\ln s/Q^2$ for equal masses    
$Q_1^2\,=\,Q_2^2\,=\,Q^2\,=\,16\,GeV^2$. Up to $Y\,\simeq 6 $    
this  corresponds to the LEP data region. For the fixed value of $\alpha_s$    
we use $\alpha_s (Q^2)\,\simeq\,0.24$.    
\FIGURE{   
\epsfig{file=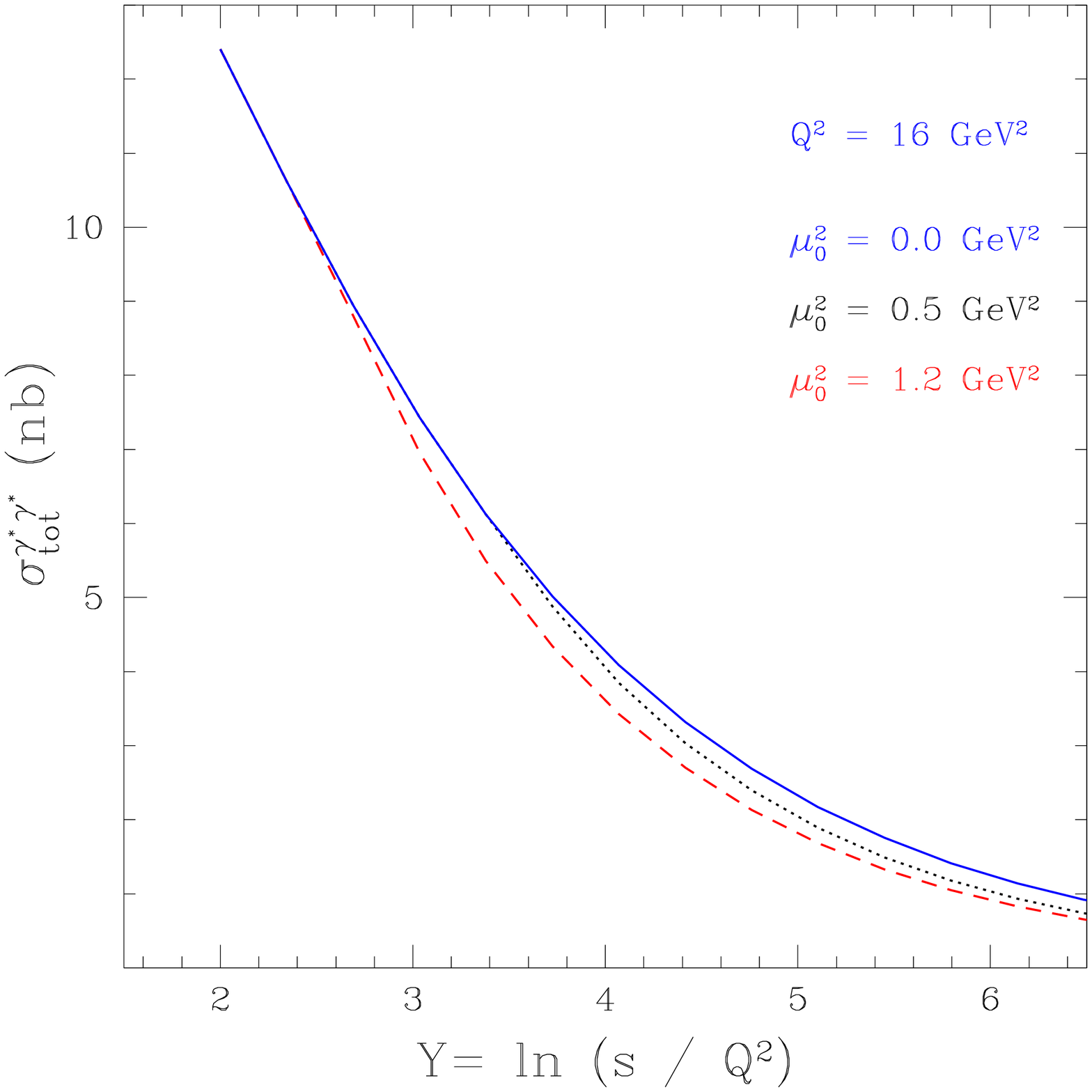,width=120mm}   
 \caption{$q\,\bar q$ contribution to     
$\sigma_{tot}^{\gamma^*\,\gamma^*}$ for various values   
of $\mu_0^2$.}   
    \label{sigfig}}   
The three curves show the    
dependence of the cross section on the nonperturbative scale $\mu_0$.    
The solid line shows the (unphysical)
 case $\mu_0^2\,=\,0$ (the region $I^+$),    
the dotted line is $\mu_0^2\,=\,0.5\,GeV^2$, and the dashed line    
is $\mu^2_0\,=\,1.2\,GeV^2$.    
The points where the different curves come together    
correspond to $s\,\mu_0^2/Q^4\,=\,1$. To the left of these points we    
have the hard domain where the perturbative QCD calculation    
is fully reliable and does not depend upon the infrared scale $\mu_0^2$.    
Note that for $\mu_0^2\,=\,0.5\,GeV^2$, almost all    
LEP data are within the hard domain. In this region we should    
expect that the secondary reggeon contribution is described by pQCD, and    
one should not add a further nonperturbative reggeon.    
    
\FIGURE{   
\epsfig{file=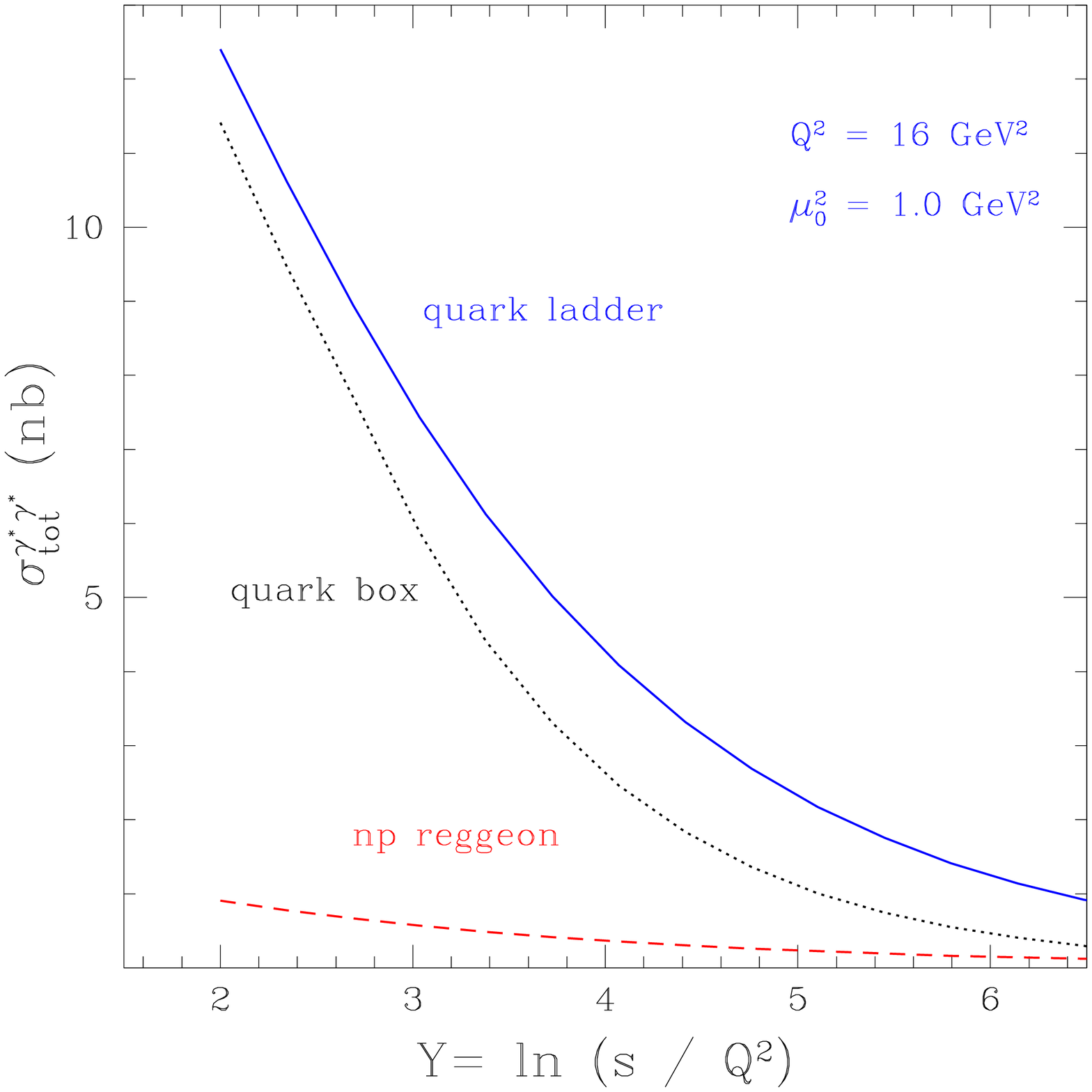,width=120mm}   
 \caption{   
Various contributions to $\sigma_{tot}^{\gamma^*\,\gamma^*}$.}   
    \label{bornfig}}

Fig. \ref{bornfig} compares the ladder resummation with the box    
diagram contribution\footnote{Only the leading logarithmic contribution is   
taken for the box diagram. The results thus obtained are somewhat larger   
compared to the ones based on the exact expression for the box.}.     
A significant enhancement is observed.    
The enhancement    
grows at higher energies and reaches a factor of ten  at $Y\,\simeq\,10$.    
For comparison we also show the nonperturbative reggeon (dashed line)    
$\sim\,s^{-\,0.45}$.  In Ref. \cite{MK}  
this contribution was added to the box    
diagram in order to fit the data. We believe that within   
the hard domain  our resummed ladder should replace the contribution of the    
phenomenological reggeon. This can be  qualitatively seen from the    
Fig. \ref{bornfig}.   
   
\section{Conclusions}

In this paper we have considered quark-antiquark exchange in    
$\gamma^*\,\gamma^*$ scattering. For the flavor nonsinglet channel we have    
obtained a closed expression for the cross section    
$\sigma^{\gamma^*\,\gamma^*}$,  valid within the double    
logarithmic accuracy of pQCD. The result depends on the four scales    
relevant for the problem $s\,\gg\,Q_1^2\,\ge\,Q_2^2\,\ge\,\mu_0^2$.    
The infrared cutoff $\mu_0$ is introduced into the analysis as the momentum    
scale below which nonperturbative physics starts.    
    
One of the central observations of this work is the role of this     
infrared cutoff $\mu_0$ in the high energy 
behavior of the total cross section.    
For large photon virtualities $Q_1^2$, $Q_2^2$, there is a region in energy    
where the internal transverse momenta lie in the hard region, and the result   
 does    
depend upon the infrared cutoff. In this region the perturbative analysis    
is reliable. When energy increases beyond the value    
$s\,=\,Q_1^2\, Q^2_2 /\mu_0^2$,    
the region of integration starts to extend into    
the infrared region; as a first step of handling this infrared dependence    
we introduce a sharp cutoff, $\mu_0^2 \,<\, k^2$,    
and the results starts to depend upon $\mu_0^2$. We find, however,  
that this dependence is weak (via a    
logarithmic prefactor only).     
    
There is a close analogy between this $\mu_0^2$  
dependence in quark-antiquark    
exchange and the diffusion in the BFKL Pomeron. In both cases, the relevant    
region of integration grows with energy; in $q\bar{q}$ exchange this growth    
is faster than in the BFKL case. As far as the $\mu_0$ dependence is    
concerned, there is also an analogy with the non-forwardness    
$t$ of the BFKL pomeron. By preventing the quark    
transverse momenta to penetrate the nonperturbative domain,    
both $\mu_0$ (in $q\bar{q}$ scattering) 
and a nonzero $t$ (in the BFKL case)     
trigger a change in the high energy asymptotics of the amplitudes:    
the pre-exponential factor in the amplitude    
changes from $1/\sqrt{\ln s}$ to $1/(\sqrt{\ln s})^3$.    
    
As a byproduct of our analysis we have derived the low-$x$ asymptotics    
of the DIS flavor nonsinglet photon structure function. 
The power dependence    
on $x$ is the same as for the proton structure function \cite{Ry}.    
    
The resummed quark ladder serves as a model for a `perturbative secondary    
reggeon'. It is    
remarkable that the resulting intercept $\omega_0\,\sim\,0.5 $ is very    
close to the one known from the high energy phenomenology. The large    
intercept is due to the fact that the leading contribution is double    
logarithmic and $\omega_0\,\sim\,\sqrt{\alpha_s}$.    
The square root dependence of $\omega_0$ on $\alpha_s$ somewhat reduces the    
uncertainty due to unknown value of $\alpha_s$.    
    
The study of the quark ladder has an obvious phenomenological motivation.    
The LEP data on $\gamma^*\,\gamma^*$ are at energies at which the quark box    
(QPM) still gives a dominant contribution to the cross section.    
We have shown that the gluon radiation leads to a 
significant enhancement of    
the quark box and hence needs to be accounted for.    
The quark box contribution dies fast with energy and is correctly expected    
to be of no importance for $\gamma^*\,\gamma^*$ scattering at a NLC.    
In contrast, the pQCD reggeon receives    
an enhancement of about a factor of ten compared to the quark box,    
and potentially can still give a noticeable correction to the dominant    
pomeron contribution.    
    
Our analysis is incomplete, since we have not yet calculated the    
flavor singlet    
quark-antiquark exchange. This step is more involved due to 
the admixture of    
$t$-channel gluons (which are in helicity states different from the    
BFKL Pomeron). The situation is similar to the flavor singlet    
contribution to the polarized structure function $g_1$ ~\cite{BER2}:    
here it is the antisymmetric $\epsilon$-tensor which, 
in the gluon t-channel,    
projects on the nonleading helicity configuration.     
The evaluation of this contribution to $\gamma^*\,\gamma^*$ is in progress    
and will be reported in a further publication.    
    
In our work, the $\gamma^*\,\gamma^*$ scattering via quark ladder exchange    
was investigated by three different methods. First we have derived and solved    
a Bethe-Salpeter-type equation for the scattering amplitude.    
We have discovered the two distinct kinematical regions described above:     
in the first region the resulting amplitude is purely perturbative and is    
not influenced by the nonperturbative scale $\mu_0$, whereas in the second    
region, the amplitude is infrared sensitive.    
The second method which was applied    
is the nonlinear IREE. Using the IREE formalism known from the literature    
and adjusting it to $\gamma^*\,\gamma^*$, we have reproduced the results    
obtained    
from the linear equation. A strong advantage of the IREE formalism lies in    
the fact that it allows to generalize to odd-signature exchange which    
includes nonladder Feynman diagrams.       
Finally, the method relying on the reggeon Green function was used.    
Using this method we successfully reproduced our result in the perturbative    
region ($\mu_0^2\,=\,0$). We do not know    
whether this method can be applied to the flavor singlet exchange.    
On the other hand, the Green's function approach allows to generalize to    
the case of nonzero momentum transfer, and it also may allow to extend    
our results to single logarithmic accuracy.    
It is claimed in Ref. \cite{Kirschner} that the Green function $G$    
(\ref{green}) is obtained within a single logarithmic accuracy, which    
is beyond the double log accuracy adopted in the present work.

\acknowledgments{We wish to thank Boris Ermolaev, Victor    
Fadin, Dima Ivanov, Roland Kirschner, Eugene Levin, Lev Lipatov,    
Misha Ryskin, Anna Stasto, and Lech Szymanowski for very    
fruitful discussions.    
    
This research was supported in part by the GIF grant $\#$    
I-620-22.14/1999. }

\appendix    
\section{Appendix}\label{sec:A}   
In this Appendix we derive the result (\ref{fi-}) for the function   
$\phi_-$. The function $\phi_-(z)$ can be found if    
(\ref{sol-}) is substituted to Eq. (\ref{BS-1}). Instead we will    
solve two simpler equations which follow from Eq. (\ref{BS-1}).    
This will be enough to uniquely determine $\phi_-$.

The first equation is the boundary condition   
\beq\label{BC}    
A^-(0,\,\eta)\,=\,A^+(L_0, \,\eta),  
\eeq   
which can be rewritten as    
\beq\label{BC1} \int_{0^+} dz \,\phi_-(z)\,e^{\eta\,z}\,=\,    
\int_{0^+} \frac{dz}{z}\,e^{\eta\,z\,+\,\lambda\,L_0/z}\,.  
\eeq    
Eq. (\ref{BC1}) implies    
\beq\label{tfi}\phi_-(z)\,=\,\frac{1}{z}\,e^{\lambda\,L_0/z}\,+\,    
\tilde\phi_-(z)\eeq    
with $\tilde\phi_-(z)$ being a function regular at the origin.

The second equation, which we are going to use, is obtained by    
differentiating Eq. (\ref{BS-1}) with respect to $\xi^\prime$. We    
get     
\beq\label{E2} \frac{\partial A^-}{\partial    
\xi^\prime}\,=\,\lambda\,\int_{\xi^\prime}^\eta d    
\bar\eta\,A^-(\xi^\prime\,,\bar\eta)  
\eeq     
which reduces to    
\beq\label{SE} \int_{0^+}\frac{d    
z}{z}\,\phi_-(z)\,e^{\xi(z\,+\,\lambda/z)}\,=\,0  
\eeq   
and has to fulfilled for any    
$\xi$. Eq. (\ref{SE}) is solved by substituting (\ref{tfi}) and    
expanding $\tilde\phi_-$ in a power series in $z$    
$$\tilde\phi_-(z)\,=\,\sum_{n=0}C_n\,z^n\,.$$    
Eq. (\ref{SE}) can be rewritten     
\beq\label{SE1} \int_{0^+}\frac{d    
z}{z}\,e^{\xi\,z\,+\,\lambda(\xi\,+\,L_0)/z}\,=\,-\,\sum_{n=0}C_n\,\int_{0^+}    
dz\,z^{n-1}\,e^{\xi(z\,+\,\lambda/z)}\,.   
\eeq     
Expanding on both sides of Eq. (\ref{SE}) the exponents in powers of $1/z$ and integrating    
over $z$ we obtain    
\beq\label{sum} \sum_{m=0}^\infty\sum_{k=0}^m    
\lambda^m\,\frac{L_0^{m-k}\,\xi^{m+k+1}}{k!\,(m-k)!\,(m+1)!}\,=\,   
-\,\sum_{n=0}^\infty    
\sum_{m=n}^\infty C_n\,\lambda^m\,\frac{\xi^{2\,m-n}}{m!\,(m-n)!} \,.   
\eeq     
The coefficients in front of a given power $\xi^p$ should be    
equal for any $p$. This leads to    
\beq\label{sum1}    
\sum_{m=(p-1)/2}^{p-1}\lambda^m\,\frac{L_0^{2\,m-p+1}}{(p-m-1)!\,   
(2\,m-p+1)!\,(m+1)!}\,=\,    
-\,\sum_{n=0}^p    
C_n\,\frac{\lambda^{(p+n)/2}}{((p+n)/2)!\,((p-n)/2)!}\,.\eeq    
Introducing a new summation index $\tilde    
m\,=\,2\,m\,-\,p\,+\,1$ casts the left hand side of (\ref{sum1})    
into the same form as the right hand side. We finally find    
\beq\label{coeff}    
C_n\,=\,-\,\frac{1}{\lambda}\frac{L_0^{n-1}}{(n-1)!}    
\,\,\,\,\,\,\,\,\,\,\,\,\,\,\,\,\,\,\,   
and\,\,\,\,\,\,\,\,\,\,\tilde\phi_-(z)\,=\,    
-\,\frac{z}{\lambda}\,e^{z\,L_0}\,.\eeq

\section{Appendix} \label{sec:B}   
    
In this appendix we compute the amplitude $T^-$ by the inverse    
Mellin transform of the expression (\ref{solIREE2}).    
Let us compute the amplitude $T^-$ as a sum of two separate terms    
$T^-=T_1\,+\,T_2$ where    
\beq\label{BT}    
T_1\,=\,\kappa\,\int_{C_{cut}}\frac{d\omega}{2\,\pi\,i}\left(\frac{s}    
{\tilde\mu^2}\right)^\omega\,    
\left[-\,\frac{f_0^2}{\omega\,-\,f_0/4\pi^2}    
\left(\frac{\tilde\mu^4}{\tilde    
Q^4}\right)^{\omega\,-\,f_0/8\pi^2}\,\right]    
\eeq    
\begin{eqnarray}\label{BT2}    
T_2&=&16 \,\pi^4\, \omega_0^2 \,\kappa \,   
\int_{C_{cut}}\frac{d\omega}{2\,\pi\,i} \left(\frac{s}    
{\tilde \mu^2}\right)^{\omega} \nonumber  \\   
&\times & \left[\,\left(\frac{\tilde Q^2}    
{\tilde \mu^2} \right)^{-\omega} \left(\frac{Q_2^2}{Q_1^2}\right)^    
{\sqrt{\omega^2 -\omega_0^2}/2}    
\frac{1} {(\omega-f_0/4\,\pi^2)}\,    
\,-\,\left(\frac{Q_1^2}{\tilde \mu^2}\right)^{-\omega}\,   
\frac{1}{\omega}\right].    
\end{eqnarray}   
As we will see below, the $\omega$-integration path $C_{cut}$   
goes around the square root branch cut     
from $-\omega_0$ to $\omega_0$.    
    
For the first term in the amplitude we have    
\beq\label{T1}    
T_1\,=\,-\,\kappa\,16\,\pi^4 \int_{C_{cut}}    
\frac{d\omega}{2 \pi i} \left(\frac{s}{\tilde    
Q^2}\right)^\omega\,\left(\frac{\tilde    
Q^2}{\tilde\mu^2}\right)^{-\sqrt{\omega^2-\omega_0^2}}    
\frac{(\omega\,-\sqrt{\omega^2\,-\,\omega_0^2})^2}    
{\sqrt{\omega^2\,-\,\omega_0^2}}.    
\eeq    
Denote by $\xi_0\,=\,\ln (s\,\tilde\mu^2/\tilde Q^4)$ and $\eta_0\,=\,\ln    
(s/\tilde\mu^2)$ and    
introduce a new complex variable    
\beq\label{ztrans}    
z\,\,=\,\, (\omega\,-\,\sqrt{\omega^2\,-\,\omega^2_0})/\omega_0\,.    
\eeq    
The $\omega$ integral then turns into a contour integral in the $z$-plane,    
and the integration path encircles the origin $z\,=\,0$. We obtain:    
\beq\label{T12}    
T_1\,=\,\kappa\,16\,\pi^4\,\omega_0^2 \int_{0^+}    
\frac{dz\,z\,}{2 \pi i}e^{\eta_0\,\omega_0\,z/2\,+\,\xi_0\,\omega_0/2\,z} \,   
=\,    
-\,\kappa\,16\,\pi^4\,\omega_0^2\,    
\frac{\xi_0}{\eta_0}\,I_2(\omega_0\,\sqrt{\xi_0\,\eta_0}),    
\eeq    
quite in agreement with the second term on the rhs of (\ref{Tamp-}).    
    
For the second part $T_2$ we have    
\beq\label{T2}    
T_2\,=\,\kappa\,\omega_0^2 16\,\pi^4\int_{C_{cut}}    
\frac{d\omega}{2 \pi i} \left(\frac{s}{    
Q^2_1}\right)^\omega\,\left[\frac{1}{\sqrt{\omega^2-\omega_0^2}}    
\left(\frac{Q_1^2}{Q_2^2}\right)^{(\omega\,-\,\sqrt{\omega^2-\omega_0^2})/2}    
\,-\,\frac{1}{\omega}\right].    
\eeq    
Denote by $\xi_1\,=\,\ln (s\,/Q_1^2)$ and $\eta_1\,=\,\ln (s/Q_2^2)$,    
and introduce $z$ as above. We obtain:    
\beq\label{T22}    
T_2\,=\,\kappa\,16\,\pi^4\,\omega_0^2\,\left[ \int_{0^+}    
\frac{dz}{2\,\pi\,i\,z}\,    
e^{\eta_1\,\omega_0\,z/2\,+\,\xi_1\,\omega_0/2\,z}\,-\,1\,\right] \,=\,    
\kappa\,16\,\pi^4\,\omega_0^2\,\left(\,    
I_0(\omega_0\,\sqrt{\xi_1\,\eta_1})\,-\,1\,\right),    
\eeq    
which agrees with the first term on the rhs of (\ref{Tamp-}).    
    
Note that the rhs of (\ref{T2}), although it looks asymmetric in    
$y_1$ and $y_2$, is completely symmetric under the the exchange    
$y_1 \leftrightarrow y_2$. This is because, in (\ref{T22}),       
we have the freedom to substitute $z \to 1/z$,    
i.e. to interchange the two momentum scales $Q_1^2$ and $Q_2^2$.    
The transformation (\ref{ztrans}) maps a single sheet $z$-plane   
to a two-sheeted $\omega$ plane. The simple change of variable:    
$z \to 1/z$    
is equivalent to replacing $\omega - \sqrt{\omega^2 - \omega_0^2}    
\to \omega + \sqrt{\omega^2 - \omega_0^2}$    
(and to moving the $\omega$ contour    
from the first to the second sheet).    
Also, the last term on the rhs of (\ref{T2}) leaves the freedom to    
multiply with $(\frac{s}{Q^2_1})^{\omega}$ or with    
$(\frac{s}{Q^2_2})^{\omega}$: in both cases, the $\omega$ integration        
gives $1$.

\end{document}